\documentclass[twocolumn,showpacs,preprintnumbers,amsmath,amssymb,floatfix,prx,superscriptaddress]{revtex4-1}
\usepackage{epstopdf}
\usepackage[latin1]{inputenc} 
\usepackage{setspace}
\usepackage{graphicx}
\usepackage{graphics}
\usepackage{subfigure}
\usepackage{dcolumn}
\usepackage{bm}
\usepackage{color}
\usepackage{SIunits}
\usepackage[normalem]{ulem}
\usepackage{url}
\usepackage[left, modulo, pagewise]{lineno}

\bibpunct{[}{]}{,}{n}{,}{,}

\begin{document}

\title{Identification of coupling mechanisms \\between ultraintense laser light and dense plasmas}

\author{L. Chopineau}	
\affiliation{LIDYL, CEA, CNRS, Universit\'e Paris-Saclay, CEA Saclay, 91 191 Gif-sur-Yvette, France\\}
\author{A. Leblanc}
\affiliation{LIDYL, CEA, CNRS, Universit\'e Paris-Saclay, CEA Saclay, 91 191 Gif-sur-Yvette, France\\}
\author{G. Blaclard} 
\affiliation{LIDYL, CEA, CNRS, Universit\'e Paris-Saclay, CEA Saclay, 91 191 Gif-sur-Yvette, France\\}
\affiliation{Lawrence Berkeley National Laboratory, Berkeley, CA 94720, USA\\}
\author{A. Denoeud}
\affiliation{LIDYL, CEA, CNRS, Universit\'e Paris-Saclay, CEA Saclay, 91 191 Gif-sur-Yvette, France\\}
\author{M. Th\'evenet}
\affiliation{Lawrence Berkeley National Laboratory, Berkeley, CA 94720, USA\\}
\author{ J-L. Vay} 
\affiliation{Lawrence Berkeley National Laboratory, Berkeley, CA 94720, USA\\}
\author{G. Bonnaud}
\affiliation{LIDYL, CEA, CNRS, Universit\'e Paris-Saclay, CEA Saclay, 91 191 Gif-sur-Yvette, France\\}
\author{ Ph. Martin}
\affiliation{LIDYL, CEA, CNRS, Universit\'e Paris-Saclay, CEA Saclay, 91 191 Gif-sur-Yvette, France\\}
\author{ H. Vincenti}
\email[]{henri.vincenti@cea.fr}
\affiliation{LIDYL, CEA, CNRS, Universit\'e Paris-Saclay, CEA Saclay, 91 191 Gif-sur-Yvette, France\\} 
\author{F. Qu\'er\'e}
\email[]{fabien.quere@cea.fr}
\affiliation{LIDYL, CEA, CNRS, Universit\'e Paris-Saclay, CEA Saclay, 91 191 Gif-sur-Yvette, France\\}

\date{\today}
\newcommand{\e}{\mbox{\boldmath$\eta$}}
\newcommand{\x}{\mbox{\boldmath$x$}}
\newcommand{\si}{\mbox{\boldmath$\xi$}}

\begin{abstract}
The interaction of intense laser beams with plasmas created on solid targets involves a rich non-linear physics. Because such dense plasmas are reflective for laser light, the coupling with the incident beam occurs within a thin layer at the interface between plasma and vacuum. One of the main paradigms used to understand this coupling, known as Brunel mechanism, is expected to be valid only for very steep plasma surfaces. Despite innumerable studies, its validity range remains uncertain, and the physics involved for smoother plasma-vacuum interfaces is unclear, especially for ultrahigh laser intensities. We report the first comprehensive experimental and numerical study of the laser-plasma coupling mechanisms as a function of the plasma interface steepness, in the relativistic interaction regime. Our results reveal a clear transition from the temporally-periodic Brunel mechanism to a chaotic dynamic associated to stochastic heating. By revealing the key signatures of these two distinct regimes on experimental observables, we provide an important landmark for the interpretation of future experiments. 
\end{abstract}

\maketitle


High-density plasmas can be created by focusing intense laser pulses on initially-solid targets. The interaction of such plasmas with laser light, investigated for several decades, involves a rich non-linear physics which is not only of fundamental interest, but also of high relevance for a wide range of applications over a large interval of laser intensities, and spanning thermonuclear fusion \cite{Fusion}, laboratory astrophysics \cite{remington2000review}, or laser-driven particle acceleration \cite{DaidoRPPion, MacchiRMP}. For most if not all applications, depositing laser energy into the plasma is essential. Due to their high density, largely in excess of the so-called critical density where the local electron plasma frequency equals the laser frequency, these plasmas however tend to reflect a large fraction of the laser light. The actual coupling with the incident light field can only occur either in the undercritical part of the density gradient at the plasma-vacuum interface, where the laser wave propagates, or within the skin depth of the overcritical plasma, where the laser wave is evanescent.

At such plasma densities, physicists initially anticipated the main mechanism of energy deposition to be collisional absorption \cite{kruer}: electron-ion collisions disrupt the regular quivering motion of the plasma electrons in the light field, statistically leading to a net kinetic energy gain from the laser field. 
Soon after the invention of lasers, experimental studies on the feasibility of laser-driven thermonuclear fusion however revealed the importance of non-collisional light absorption mechanisms, coming into play for moderate laser intensities ($I \lambda^2 \gtrsim 10^{13}$ $W \mu m^2/cm^2$) \cite{Labaune}. For interactions at the surface of dense plasmas, these processes are expected to be most relevant when the laser beam impinges the target at oblique incidence (angle of incidence $\theta_i \neq 0$) and in $p$-polarization, such that an electric field component efficiently drives electron motion along the normal to the target surface. Among these so-called 'anomalous absorption' mechanisms, the resonant excitation and subsequent damping of collective electronic plasma waves at the critical plasma density has been the first key process to be identified both theoretically and experimentally, and is commonly known as resonant absorption \cite{RA1,RA2,RA3,RA4}.  

Since the coupling with laser light occurs at the interface of the plasma with vacuum, the characteristic spatial length $L$ of the plasma density gradient across this interface is a crucial parameter. This density gradient is generally not step-like, due to the unavoidable expansion of the plasma into vacuum, either during the main laser pulse driving the interaction, or even before this pulse when laser pre-pulses are present (either accidentally, or voluntarily introduced). In an influential paper from 1987, Brunel predicted a transition from resonant absorption to a new coupling mechanism, that he ironically called 'not-so-resonant, resonant absorption' \cite{brunel}, depending on the value of $L$. He anticipated this mechanism to come into play when the laser intensity becomes so strong that the quivering amplitude of the plasma electrons in the field along the surface normal gets larger than $L$, such that the plasma-vacuum interface can be modeled as step-like. 

\begin{figure}[t]
\centering \includegraphics[width=1\linewidth]{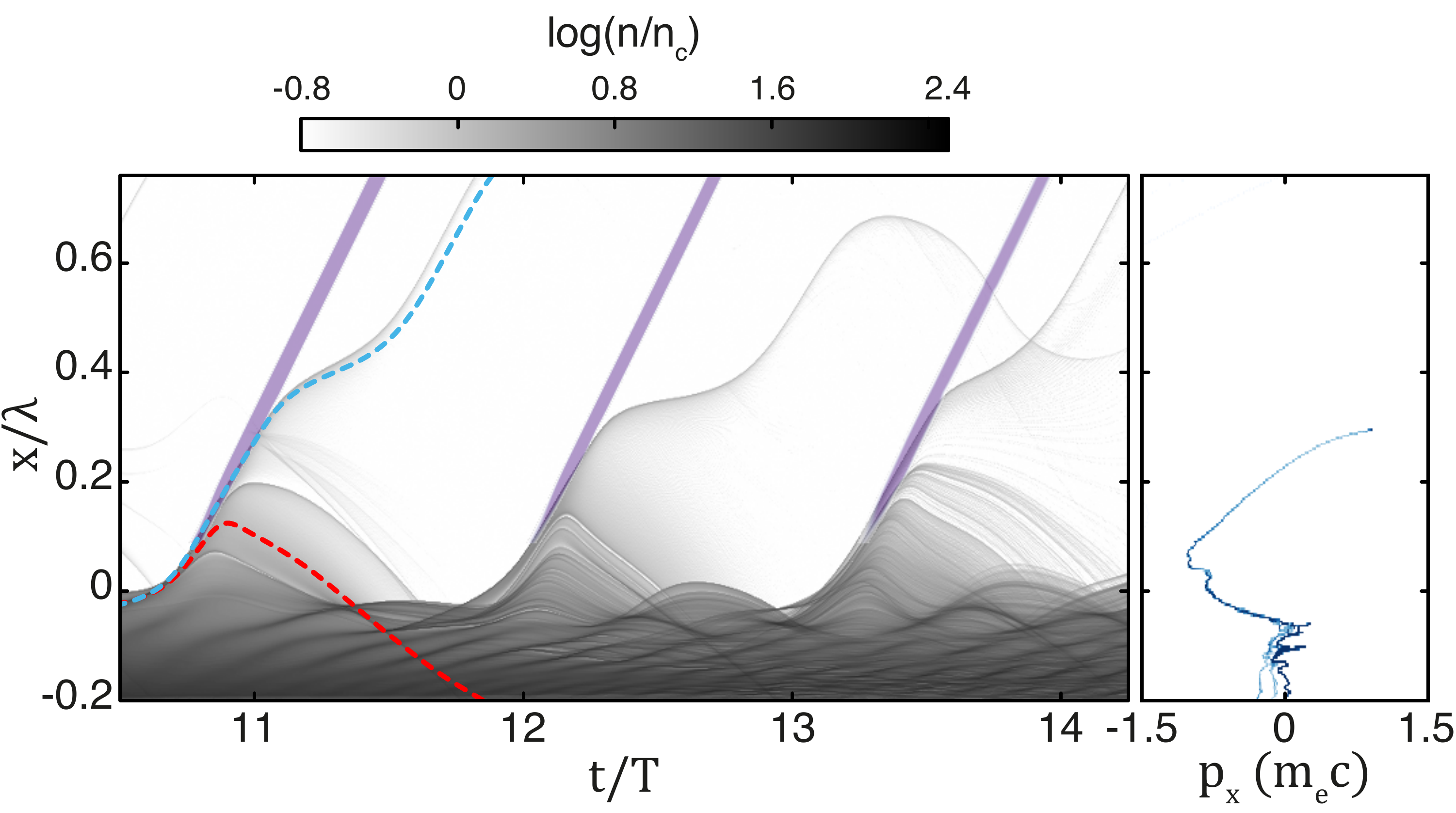}
\vskip -0.25cm 
\caption{\textbf{Temporal dynamics of a dense plasma exposed to an ultraintense laser field in the Brunel regime.} This graph displays results from a Particle-In-Cell simulation performed for $a_0=2$ ($I=8.5\times10^{18}$ $W/cm^2$ for $\lambda=800$ nm), $\theta_i=55^o$, and a density gradient scale length $L=\lambda/10$ (with $\lambda$ the laser wavelength). The gray-scale color map shows the temporal evolution, during three laser optical periods $T$, of the plasma electron density around the target surface, while the purple color scale shows the attosecond light pulses emitted by this plasma (harmonics 8 to 15). Two representative trajectories for the particles forming the expelled electron beam (blue color) and the 'recolliding' electron flux (red color) are also displayed. The right panel shows a typical distribution of electrons in the $x-p_x$ phase space ($x$ spatial coordinate along the target normal) at time $t/T=12.7$.} 

\vskip -0.5cm
\label{fig-PIC1D-Brunel}
\end{figure}

This simple and intuitive mechanism, now known as Brunel absorption or vacuum heating, is qualitatively analogous to the intensively studied three-step model of atomic and molecular strong-field physics, where an intense laser field ($I \lambda^2 \gtrsim 10^{13}$ $W \mu m^2/cm^2$) drives the recollision of ionized electrons with their parent ions \cite{corkum}. Here, in each optical cycle,  electrons at the target surface are dragged out of the plasma into vacuum when the component of the laser $E$-field normal to the surface points to the target (see simulation results in Fig.\ref{fig-PIC1D-Brunel}). Later in the cycle, when this driving field changes sign, some of these expelled electrons are pushed back toward the 'parent plasma': as they penetrate into this dense plasma, they escape the influence of the laser field due to plasma screening, and propagate ballistically into the target (red trajectory in Fig.\ref{fig-PIC1D-Brunel}). The initial model by Brunel for a step-like surface of perfectly-conducting plasma focused on this returning electron population. However, numerical simulations show that for smoother interfaces and/or higher laser intensities, another fraction of the electrons escapes into vacuum (blue trajectory in Fig.\ref{fig-PIC1D-Brunel}), typically in the form of periodic attosecond bunches. In both cases, the fast electrons resulting from this sub-optical-cycle dynamics carry away energy acquired from the laser: for convenience, the terms 'Brunel electrons' or 'Brunel absorption' used in this paper will encompass these two populations.

A few years later, with the development of high-power femtosecond lasers, Brunel mechanism appeared as an ideal 'toy-model' to understand the interaction of dense plasmas with these ultrashort pulses. First, their ultrahigh intensities result in large electron quivering amplitudes. Second, these pulses are so short that plasma expansion during the interaction is very limited, potentially leading to sharp density gradients at the plasma surface, i.e. small values of  $L$. However, from an experimental point of view, reaching this regime turned out to be much more challenging than expected. The key difficulty arose from the unavoidable light pedestal ahead of ultrashort laser pulses \cite{Nantel}: this pedestal, if too intense, leads to the premature creation and expansion of the plasma, and thus to long and largely uncontrolled density gradients at the plasma-vacuum interface when the main laser pulse hits the target, making the Brunel regime inaccessible. More generally, this major issue has considerably complicated the interpretation of most early experiments on the interaction of intense ultrashort lasers with dense plasmas.

It took more than an additional decade to find methods to efficiently reduce the light pedestal accompanying ultrashort laser pulses \cite{ITATANI, kapteyn_mp, doumy, dromey_mp}, and thus obtain temporal contrasts that at last made extremely sharp plasma surfaces accessible and compatible with ultrahigh laser intensities \cite{levy, article_NP}. Nowadays, Brunel mechanism is most likely at play in experiments performed on solid targets with ultraintense laser pulses of suitably high-contrast. Yet, direct experimental evidence is still elusive, and its range of validity is not precisely known so far. Furthermore, following the historical development of this topic, the 'common wisdom' tends to be that when $L$ is increased, a transition from Brunel to resonant absorption should at some point occur \cite{gibbon1992collisionless, gibbon1999, FabienJPB2010}, but no clear experimental evidence of this transition has been reported yet.

A broad range of topical experiments are now performed worldwide on the interaction of ultraintense laser pulses ($I \lambda^2>10^{18}$ $W \mu m^2/cm^2$) with dense plasmas, driven by applications such as laser-driven ion  \cite{beg1997study, wilks2001energetic, DaidoRPPion, MacchiRMP} and electron acceleration \cite{wharton1998experimental, bastiani1999hot, cai2003experimental, mordovanakis2009quasimonoenergetic, brandl2009directed, wang2010angular, tian2012electron, VLA}, or the generation of intense harmonics and/or attosecond light pulses \cite{FabienJPB2010, teubner_RMP}. Clearly identifying the laser-plasma coupling mechanisms at play in this interaction regime, and determining the range of physical parameters where they are relevant, is essential for the proper understanding of such experiments. This is what we achieve in this article, by focusing ultraintense femtosecond laser pulses on a dense plasma with a sharp, controlled and measured density gradient scale length $L$, which we systematically vary from $L \ll \lambda$ to $L\approx\lambda$ (with $\lambda$ the laser wavelength). We show how performing and correlating measurements of the high-order harmonics and relativistic electrons emerging from the target provide clear signatures of these couplings mechanisms, and relate these observations to the underlying physics through an advanced analysis of 2D and 3D Particle-In-Cell (PIC) simulations, solving the coupled Vlasov-Maxwell equation system.

\begin{figure*}[t]
\centering \includegraphics[width=1\linewidth]{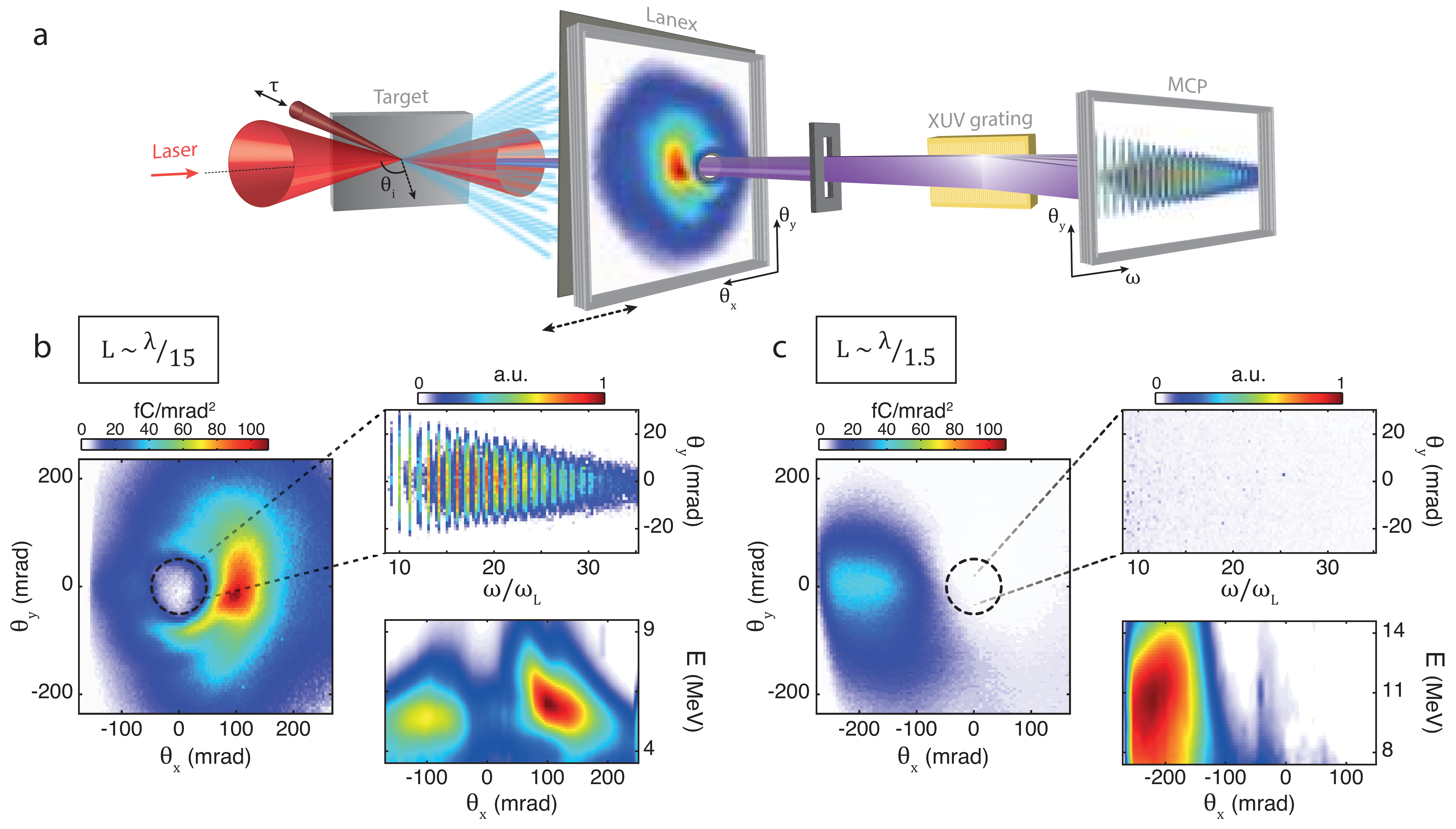}
\vskip -0.25cm 
\caption{\textbf{Principle of the experiment and main experimental findings.} Panel a shows a sketch of the experiment. The target is impinged by a controlled prepulse, followed by the main pulse after an adjustable delay $\tau$. Two of the main diagnostics are displayed, the \textsc{lanex} screen for the measurement of the spatial profile of the high-energy electron beam, and the angularly-resolved \textsc{xuv} spectrometer. These two diagnostics can be used either separately, or simultaneously when small holes are made in the electron detection assembly, as shown in the figure. They can also be replaced by an angularly-resolved electron spectrometer. The main experimental findings for a $p$-polarized laser field are summarized in the two lower panels b and c: left images, angular emission pattern of relativistic electrons; bottom right images, angularly-resolved energy spectrum of electrons in the incidence plane ($\theta_y=0$); top right images, angularly-resolved harmonic spectrum. These illustrate the major changes occurring on these three observables as the density gradient scale length is increased from $L \ll \lambda$ to $L \sim \lambda$ ($\theta_i=55^o, a_0=3.5$, $\tau=1$ $ps$ for panel b, leading to $L=\lambda/15$, $\tau=10$ $ps$ for panel c, leading to $L=\lambda/1.5$). These very contrasted features constitute signatures of the different underlying laser-plasma coupling mechanisms.} 

\vskip -0.5cm
\label{fig-Exp-Principle}
\end{figure*}
This comprehensive study shows that Brunel mechanism is indeed the relevant physical process for sharp enough plasma-vacuum interfaces. As expected, a transition occurs to a different mechanism when the density gradient scale length $L$ is increased. Measurements of this transition as a function of the laser incidence angle provide confirmation of Brunel's transition criteria based on the comparison of the electron quivering amplitude with the typical spatial extent of the interface. However, we establish that in the regime of ultrahigh laser intensities considered here, resonant absorption plays no significant role in the regime of large $L$ ($L\approx\lambda$). The coupling is rather dominated by another kinetic mechanism, so far known as stochastic heating, in which collective plasma effects play little role: as initially suggested in Ref. \cite{SentokuAPB}, electrons in the underdense part of the density gradient gain energy in the interference pattern resulting from the superposition of the incident laser field with the field reflected by the overdense part of the plasma. It has been established theoretically that at the laser intensities considered here, electron dynamics in such an interference pattern is not integrable, gets chaotic, and can lead to high energy transfer from the laser wave to the electron population \cite{ZMS2002}. 

In this paper, the amplitude of the incident laser field, which determines the intensity on target, is characterized by the dimensionless potential vector at the peak of the pulse, $a_0=e E_0/m c \omega=\lambda[\mu m]  (I[W/cm^2]/1.37\times10^{18})^{1/2}$, with $c$ the speed of light, $e$ the elementary charge, $m$ the electron mass, $\omega$ the laser frequency, and $E_0$ the amplitude of the laser electric field. All experiments and simulations presented here have been performed with $a_0>1$, which corresponds to the interaction regime where relativistic effects play an important role on electron motion. We define $n_c$ as the critical plasma density associated to the laser frequency $\omega$ ($n_c=m \epsilon_0 \omega^2/e^2=1.74 \times 10^{21}$ $cm^{-3}$ for 800 nm laser light).

\section{Description of the experiment}

A sketch of the experiment is presented in the upper part of Fig.\ref{fig-Exp-Principle}. A high-power femtosecond laser beam is focused on a silica target, which it fully ionizes on a thin surface layer, thus producing a dense plasma (maximum plasma density $n_0 \simeq 6.10^{23}$ $cm^{-3}$, i.e $400 n_c$ for 800 nm light). 

We use the UHI100 laser at CEA Saclay, a commercial system delivering 20 to 25 fs pulses with a peak power of 100 TW. After correction of its wavefront by an adaptive optic system, the beam is focused by an off-axis parabolic mirror with a $f$-number of $f/6$, leading to a focal spot of 5 $\mu m$ diameter (FWHM in intensity) and to an estimated peak intensity of $2.10^{19}$ $W/cm^2$ ($a_0 \approx 3.5$) on target. By default, the laser beam is $p$-polarized on target, but the polarization can be switched to $s$ by inserting a thin zero-order mica half wave-plate in the beam.  

The first key aspect of the experiment is that it was carried out with high degree of control and an accurate knowledge of the plasma density gradient scale length $L$ at the target surface, which is a prerequisite for the study of the laser-plasma coupling mechanisms. This implies the use of laser pulses of very high temporal contrast, so that premature creation of the plasma on target is avoided: this is achieved thanks to a double plasma mirror system placed before the main experimental chamber \cite{levy}, which increases the contrast by about 4 decades, from $ \gtrsim 10^{9}$ to $\gtrsim 10^{13}$ on a $\gtrsim100$ $ps$ time scale. This ultrahigh temporal contrast is of paramount importance for all experiments presented here. 
Starting from the very steep density gradient allowed by this ultrahigh temporal contrast, $L$ is then varied in a controlled way, thanks to the introduction of a 'weak' prepulse (fluence $\approx 1$ $kJ/cm^2$) at an adjustable delay $\tau$ before the main pulse ($0\leq \tau \leq 15$ $ps$), produced from an edge of the main beam using the optical layout described in Ref.\cite{KahalyGradient}. This prepulse is strong enough to ionize the target and initiate plasma expansion, at a typical velocity in the 40 to 60 $nm/ps$ range. For all experimental conditions considered herein, and in particular for all incidence angles, we systematically measured $L(\tau)$ using the recently-introduced technique of Spatial Domain Interferometry  \cite{Bocoum:15} (SDI, see supplementary material).

The second key aspect of the experiment is the combination of diagnostics that were implemented to study the interaction. We concentrated on two types of observables: the relativistic electron beam emitted by the target towards vacuum, and the beam of high-order harmonics generated around the specular reflection direction.

Two diagnostics were used for the electron beam. First, a \textsc{lanex} screen, placed behind a 13 $\mu m$ thick aluminum foil (to eliminate laser light and its harmonics) and a 2 mm thick glass plate (to filter out low-energy electrons), and which fluorescence was imaged on a CCD camera, provided the spatial profile of the emission of electrons with energies higher than 1 MeV, at a distance of $\approx10$ $cm$ from the target (left images in Fig.\ref{fig-Exp-Principle}b-c). Second, we designed a new type of magnetic spectrometer for relativistic electrons (see supplementary material), which provided, for every laser shot, the \textsl{angularly-resolved} kinetic energy spectrum of electrons, in the incidence plane (i.e. for $\theta_y=0$). This $(\theta_x,E)$ distribution, with $\theta_x$ the angle in the plane of incidence and $E$ the electron kinetic energy, was measured with a very large angular acceptance of $\Delta \theta_x = 500$ $mrad$ around the specular reflection direction ($\theta_x=0$) (bottom right images in Fig.\ref{fig-Exp-Principle}b-c).

The harmonic beam was characterized using an angularly-resoled XUV spectrometer \cite{KahalyGradient}, with an angular acceptance of $200$ $mrad$ around the specular direction (top right images in Fig.\ref{fig-Exp-Principle}b-c). The harmonic spectrum and the electron beam spatial profile were initially measured on the same laser shots, thanks to small holes in the aluminum foil, glass plate and \textsc{lanex} screen that let the harmonic beam go through (Fig.\ref{fig-Exp-Principle}). However, we observed an excellent shot-to-shot reproducibility of the experimental results, so that these multiple diagnostics were finally implemented on different laser shots performed under identical interaction conditions.  

A simple additional diagnostic, implemented on separate laser shots, consisted in measuring the spatial profile of the laser beam reflected by the target, by inserting a frosted glass plate in this beam  20 to 30 cm after the target, and measuring the image of the laser light scattered by this plate on a camera placed behind a bandpass filter centered on the fundamental laser frequency. This can be exploited to determine the plasma reflectivity for the incident laser beam.

\section{Experimental results}
\label{Section:exp resu}

The lower part of Fig.\ref{fig-Exp-Principle} summarizes the main finding of the experiment for a $p$ polarization of the incident laser, by presenting the electron beam angular profiles and $(\theta_x,E)$ distributions, as well as the angularly-resolved harmonic spectra, measured for two different density gradient scale lengths $L$, $L_1 \ll \lambda$ and $L_2\approx \lambda$: as $L$ is increased from $L_1$ to $L_2$, the measured signals for all these observables radically change. Three main differences are observed: (i) When $L \ll \lambda$, the electron emission is predominantly peaked at $\theta_x\approx 100$ $mrad$, i.e. close to the direction of laser specular reflection ($\theta_x=0$ $mrad$), with a slight shift towards the target normal. As $L$ is increased, it then switches to $\theta_x \approx -200$ $mrad$, a direction between specular direction and the tangent to the target surface ($\theta_x=-600$ $mrad$), and simultaneously slightly broadens angularly. (ii) Electrons reach energies about twice higher for large $L$ (spectral peak around 10 MeV), with a $(\theta_x, E)$ distribution that significantly changes. In the short gradient regime, a clear correlation is observed between emission angle $\theta_x$ and electron energy $E$, especially in the most intense part of the distribution ($0\leq \theta_x\leq 200$ $mrad$): the electron energy increases as one gets closer to the specular direction. In contrast, in the long gradient regime, the electron spectrum hardly varies angularly, i.e. no significant correlation is observed on this $(\theta_x, E)$ distribution. (iii) Harmonic emission is clearly observed for small $L$, but it drops below the experimental detection threshold for large $L$. 

\begin{figure*}[t]
\centering \includegraphics [width=\linewidth]{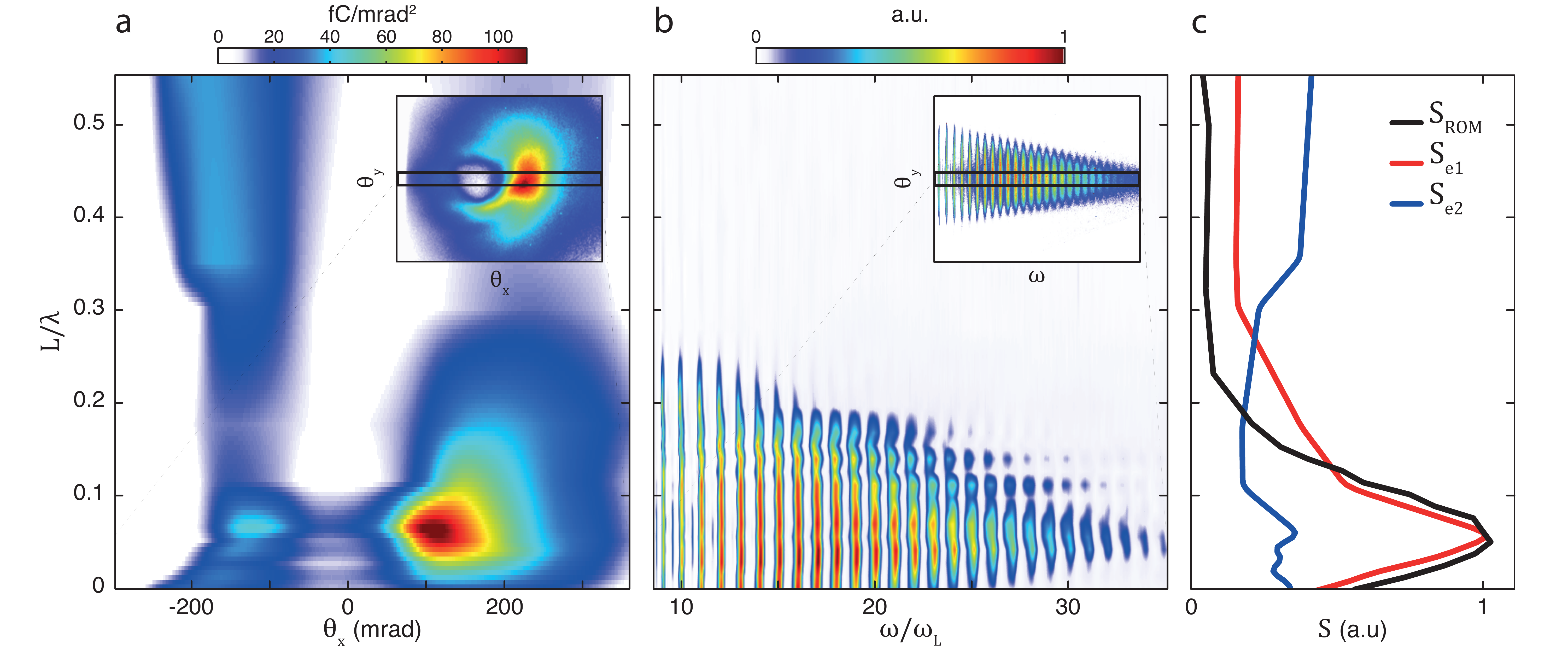}
\vskip -0.25cm 
\caption{\textbf{Evolution of the experimental observables with the density gradient scale length.} The angular profile of the relativistic electron beam in the incidence plane (panel a), and the emitted harmonic spectrum (panel b) are plotted as a function of $L$, for a $p$-polarized laser field. The experimental parameters are the same as in Fig.\ref{fig-Exp-Principle}. The insets show how these quantities are related to the images of Fig.\ref{fig-Exp-Principle} (note that there is no measured signal below these insets). Panel c shows different curves derived from these datasets: the harmonic signal integrated from the $20^{th}$ to the $25^{th}$ order (signal $S_{ROM}$), and the electron signals on the right ($50$  $mrad$ $\leq \theta_x \leq$ $150$ $mrad$, signal $S_{e1}$) and left sides ($-250$  $mrad$ $\leq \theta_x \leq -150$ $mrad$, signal $S_{e2}$) of the specular direction, are plotted as a function of $L$. These curves show the transition between the two regimes highlighted in Fig.\ref{fig-Exp-Principle}, and reveal the quantitative correlation between the harmonic signal and the relativistic electrons emission for short gradient scale lengths.   } 

\vskip -0.5cm
\label{fig-Exp-Resu}
\end{figure*}

\begin{figure}[t]
\centering \includegraphics[width=\linewidth]{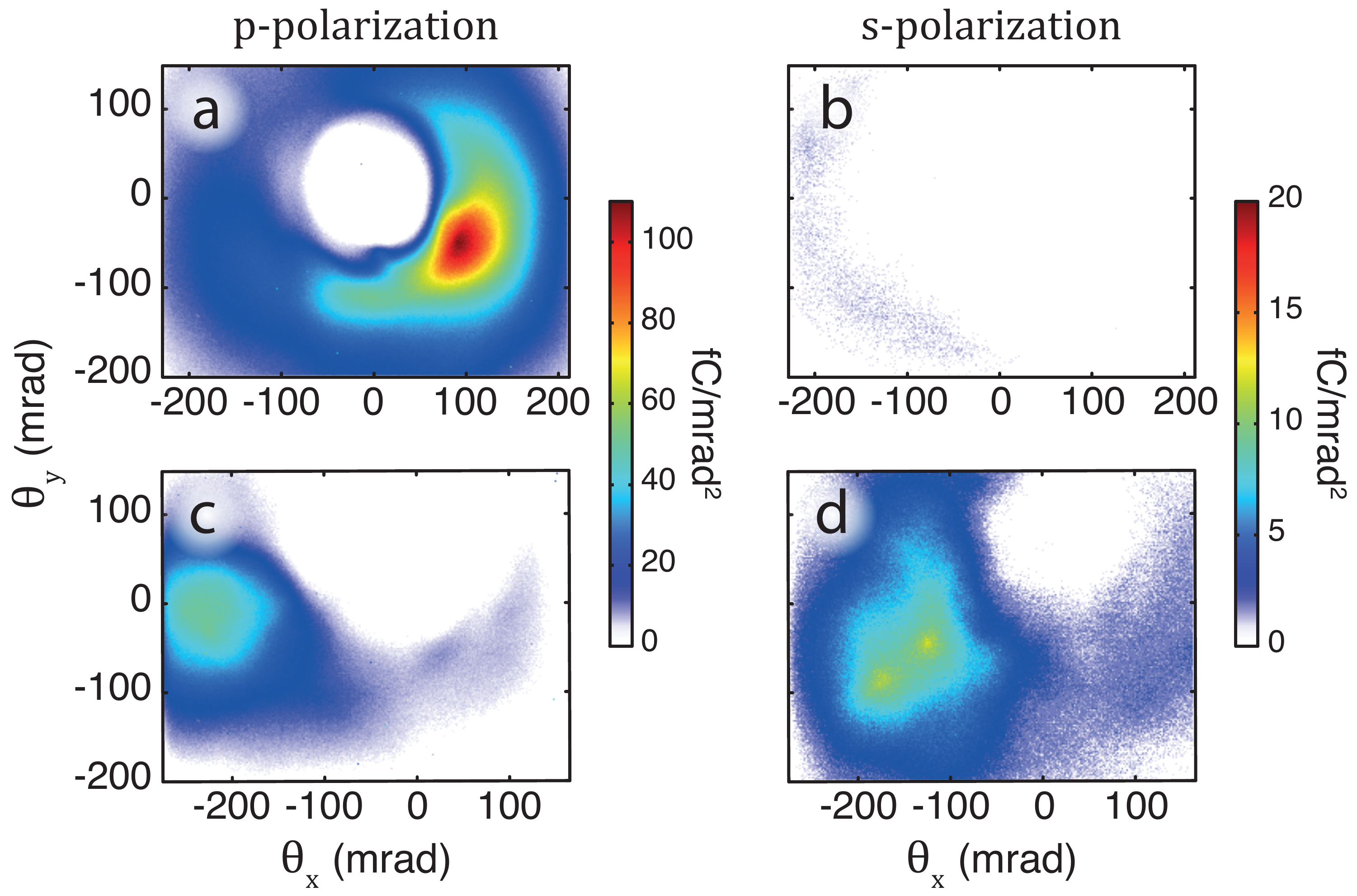}
\vskip -0.25cm 
\caption{\textbf{Effect of the laser polarization direction on relativistic electron emission.} Panel a and b show the angular profiles of the relativistic electron emission for a gradient scale length $L_1=\lambda/15$, respectively for $p$ and $s$ polarizations of the incident laser. Panels c and d show these electron angular distributions, now measured for a longer gradient $L_2=\lambda/1.5$, i.e. beyond the transition observed in Fig.\ref{fig-Exp-Resu}. Note the different color scales used in panels a-c versus b-d. } 

\vskip -0.5cm
\label{fig-Exp-PS}
\end{figure}

\begin{figure}[t]
\centering \includegraphics[width=\linewidth]{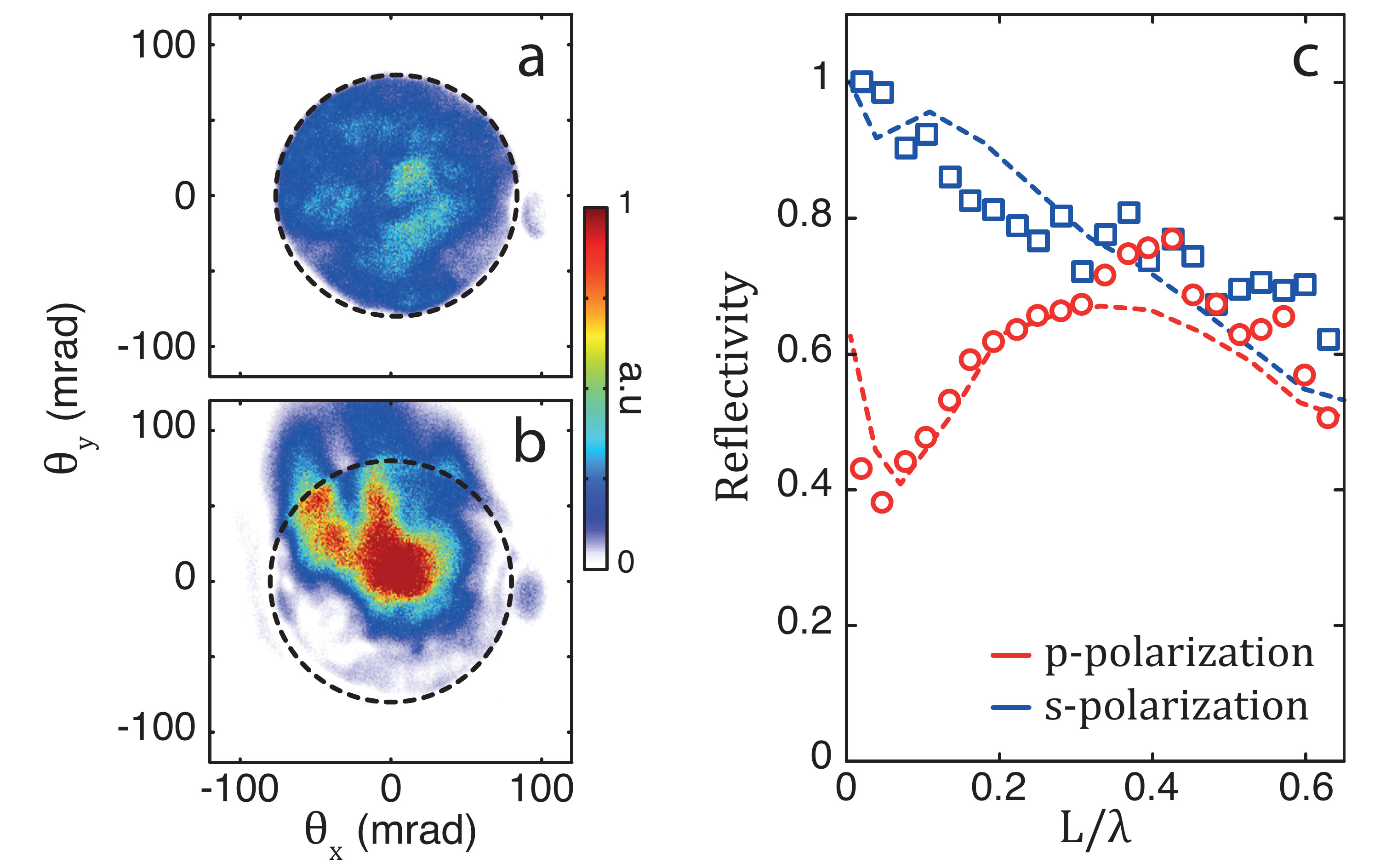}
\vskip -0.25cm 
\caption{\textbf{Reflected fundamental beam and evolution of the plasma reflectivity}. Using a scattering screen, the spatial intensity profile of the laser beam reflected by the target can be measured. The two images on the left show typical beam profiles respectively obtained in the short (upper image, $L_1=\lambda/15$) and long (lower image, $L_2=\lambda/1.5$) density gradient regimes, for $p$-polarization of the incident laser. The black dashed circles indicate the initial divergence of the top-hat laser beam, before its interaction with the target. From the spatial integration of these images, the reflectivity of the plasma for the fundamental laser frequency can be determined, and is plotted in panel c as a function of $L$ for both $s$ and $p$ polarizations (squares and circles). The lines show the corresponding results of 2D Particle-In-Cell simulations (see section \ref{Energy}). }

\vskip -0.5cm
\label{fig-Reflect}
\end{figure}

The details of the transition between these two regimes are presented in Fig.\ref{fig-Exp-Resu}, which displays the evolution with $L$ of the electron beam angular profile in the incidence plane, and of the harmonic spectrum. The most important point is the \textit{quantitative} correlation, observed at short gradients, between the emission of relativistic electrons and the harmonic signal (see curves in Fig.\ref{fig-Exp-Resu}c). As $L$ is gradually increased, the electron signal around $\theta_x=100$ $mrad$ and the harmonic signal reach a common optimum around $L=\lambda/15$, and then both quickly decrease. The electron signal on the other side of the specular direction then grows, but is not associated with any harmonic signal. The transition between these two regimes occurs around $L \approx \lambda/5$.

Another important difference between these two interaction regimes is the dependence of the observables on laser polarization direction, illustrated in Fig.\ref{fig-Exp-PS}. In the short gradient regime, the electron and harmonic signals are totally suppressed when the polarization is switched from $p$ to $s$. By contrast, for longer gradients, the electron signal is still observed for $s$-polarization, although it gets about five times weaker. 

These observations on the electron and harmonic beams clearly point to a complete change in the coupling mechanism between the laser field and the plasma, which we will analyze in the rest of this article. In the following, we will refer to these two interaction regimes as the short-gradient and long-gradient regimes for convenience. 

\begin{figure*}[t]
\centering \includegraphics[width=1\linewidth]{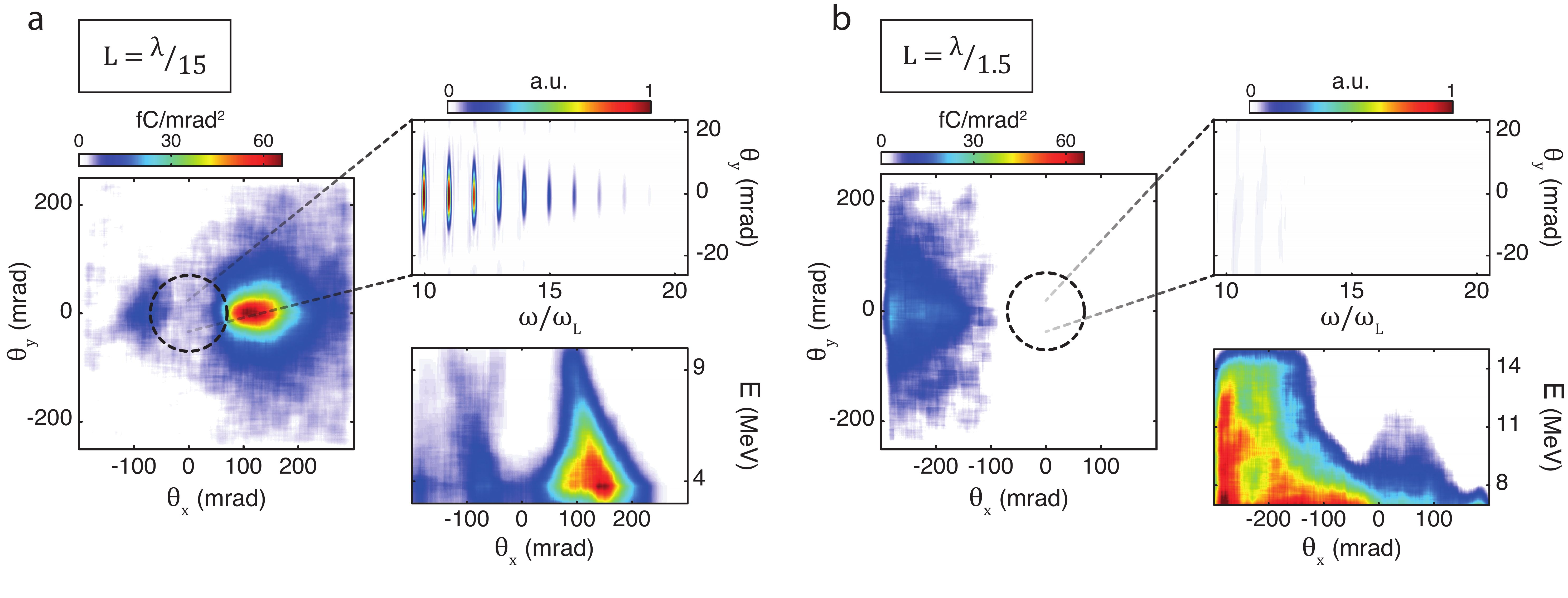}
\vskip -0.25cm 
\caption{\textbf{3D PIC simulations of the laser-plasma interaction for two different density gradient scale lengths.} The physical conditions of these simulations are matched to the estimated experimental conditions of the shots shown in the lower part of Fig.\ref{fig-Exp-Principle}: panel a corresponds to a density gradient $L_1=\lambda/15$ and panel b to $L_2=\lambda/1.5$, while all other physical parameters remain the same. From these simulations, we extract the same observables as those measured in the experiment: the angular profile of the high-energy electron beam expelled in vacuum (left image in each panel), the $(\theta_x, E)$ distribution of these electrons (bottom right image in each panel), and the angularly-resolved harmonic spectrum (top right image in each panel). The key features are the same as those observed on the experimental results (compare with the lower panels of Fig.\ref{fig-Exp-Principle}).} 

\vskip -0.5cm
\label{fig-PIC3D}
\end{figure*}
%

This transition also has consequences on even simpler observables: Fig.\ref{fig-Reflect} (left panels) displays the spatial intensity profiles of the reflected laser beam, measured on a scattering screen in these two distinct coupling regimes. In the short gradient regime, a smooth beam is observed, which is almost unaltered compared to the incident laser beam: this is the so-called plasma mirror regime \cite{article_NP}, where the plasma acts as a usual high-quality mirror, specularly reflecting the fundamental frequency, despite the ultrahigh intensity on target. By contrast, in the long gradient regime, the beam profile is strongly perturbed and starts exhibiting spatial structures that were not present on the incident beam. The term plasma mirror is thus no longer appropriate to this regime, although the laser field still interacts with a dense -and hence reflective- plasma. Experimentally, the spatial profile of the reflected laser beam might thus also be used as an alternative and very simple signature of the transition in the laser-plasma interaction.
 By spatially integrating these images, the variation of the plasma reflectivity at the fundamental laser frequency as a function of $L$ can be determined, and is displayed in the right panel of Fig.\ref{fig-Reflect}, for both $p$ and $s$ polarizations of the incident laser field. 

All these measurements have been repeated over four different experimental campaigns on the UHI100 experimental facility over the last four years, and all effects described above have been observed to be very reproducible and robust. 

\section{3D particle-In-Cell simulations}

To interpret these experimental observations, we will now turn to Particle-In-Cell (PIC) simulations of the laser-plasma interaction. This requires ensuring that (i) these PIC simulations are performed in the actual physical conditions of the experiment, (ii) they are reliable and properly reproduce the key experimental findings. To check these two critical points, we first carry out full 3D simulations of the interaction, so that we can directly confront the numerical and experimental results -especially the full angular pattern of the electron emission, which is only accessible by 3D simulations. 

All simulations reported herein have been performed using the recently-developed WARP+PXR code \cite{WARPweb, PICSAR, VAY, VINCENTI2016, VINCENTI2017}. The specificity of this code is the use of a massively-parallel high-order spectral solver for Maxwell's equations \cite{VINCENTI201822}, which greatly reduces numerical dispersion of electromagnetic waves as well as numerical noise. This ensures convergence of the simulations for larger spatial and temporal mesh steps than in most other PIC codes, and thus makes physically-realistic and reliable 3D simulations of the interaction with dense plasmas computationally tractable \cite{VINCENTI201822,VincentiPhysRevE}. Each 3D simulations reported here required 6.3 millions computation hours on a massively-parallel machine  \cite{MIRA, CORI}. More detailed information on the numerical parameters of these simulations is provided in the supplementary material. 

We performed two 3D simulations, for the same physical conditions as in the experiments of Fig.\ref{fig-Exp-Principle}b and c, corresponding to fixed laser parameters but different density gradients scale lengths, $L_1 \ll \lambda$ and $L_2\approx \lambda$. From these simulations, we extracted the exact same observables as those measured in the experiment: the calculated angular profiles and angle-energy distributions of the emitted electron beam, as well as the angularly-resolved harmonic spectra, are displayed in Fig.\ref{fig-PIC3D}. Comparison with the lower panels of Fig.\ref{fig-Exp-Principle} shows that these simulations very well reproduce the two distinct interaction regimes observed experimentally for all these observables.

These 3D benchmark simulations clearly demonstrate both the reliability of the PIC simulations, as well as our excellent control of the interaction conditions in the experiment, which ensures a relevant choice of the physical parameters used in the theoretical study. In the next section,  we will further exploit simulations performed with this WARP+PXR code to get detailed insight into the physical processes underlying these distinct interaction regimes. To this end, we will rely on more tractable 2D simulations \cite{NotePIC}. Detailed information on the numerical parameters of these simulations is also provided in the supplementary material.  

\section{Physical analysis and discussion}

\begin{figure*}[t]
\centering \includegraphics[width=\linewidth]{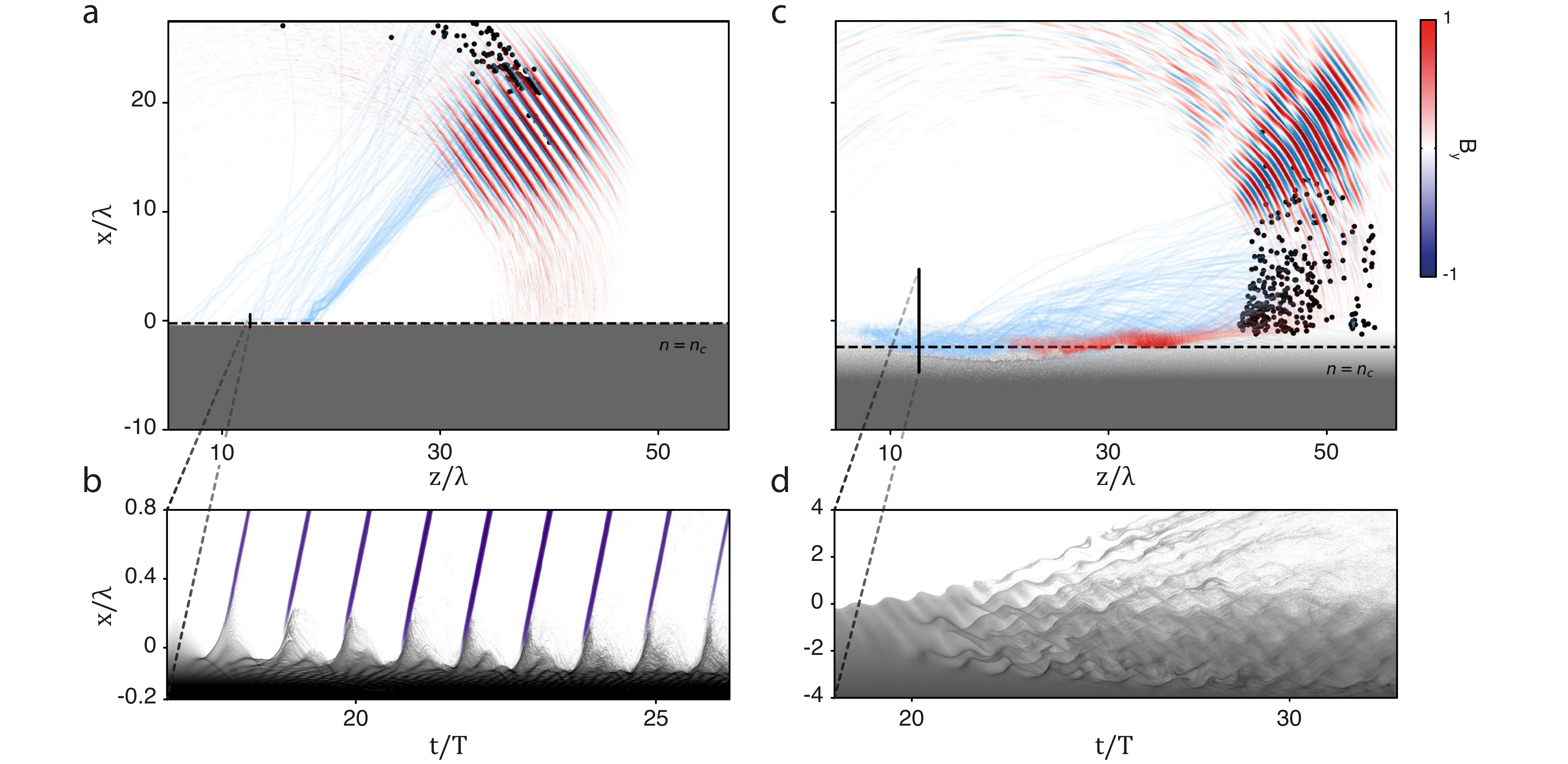}
\vskip -0.25cm 
\caption{\textbf{2D PIC simulations in the two distinct regimes of laser-plasma coupling.} These data are obtained from 2D PIC simulations with different density gradients $L$ ($\lambda/15$ for a and b, and $\lambda/1.5$ for c and d), while all other physical parameters remain the same ($a_0=3.5, \theta_i=55^o$). The two upper panels display the complete trajectories of a selected set of expelled high-energy test electrons (orange lines), together with the total $y$-component of the total B-field (blue to red color map) at a given time after the laser-plasma interaction. The plasma density profile at the end of the interaction is indicated in gray in log scale. The lower panels shows the temporal evolution of the plasma electron density (gray-scale color map, in log scale), spatially-resolved along the normal to the target surface, at the center of the focal spot. The emitted attosecond pulses are superimposed to this density map in purple. They are clearly visible in panel b, but are too weak to be observed in panel d. } 
\vskip -0.5cm
\label{fig_ejection}
\end{figure*}

The starting point for our analysis of the experimental results is the joint measurement of the relativistic electron and harmonic signals, which provides information on the temporal structure of the electron emission by the plasma (section \ref{e-HHG}). We then discuss the spatial structure of the observed electron beams (section \ref{e-spatial}). In order to understand the electron heating and ejection mechanism in the long gradient regime, we turn to a detailed analysis of PIC simulations, presented in section \ref{mechanism-long}. With the physical insight provided by this analysis, we finally discuss the influence of the laser polarization (Fig.\ref{fig-Exp-PS}) and the evolution with $L$ of the plasma reflectivity at the fundamental laser frequency (Fig.\ref{fig-Reflect}) in section \ref{Energy}. 

\subsection{Temporal structure of the electron emission} \label{e-HHG}

The clear correlation observed between the high-energy electron signal and the harmonic signal for short gradients (Fig.\ref{fig-Exp-Resu}c) suggests that in this regime, the relativistic electrons are involved in the harmonic emission. Then, the fact that a highly-contrasted harmonic comb is produced would be an indication that this electron emission is periodic in time, being locked to the driving laser field.  

To support this tentative interpretation, we consider the PIC simulations of Fig.\ref{fig-PIC1D-Brunel} and Fig.\ref{fig_ejection}b: they show that temporal periodicity is indeed a characteristic of the Brunel mechanism. Electron emission occurs in the form of bunches that are initially extremely short (in the attosecond range), emitted once every optical period. For $a_0>1$, these electrons reach relativistic velocities when they escape the plasma, and thus induce a Doppler effect on the reflected laser field: this results in the generation of a train of attosecond light pulses,  spaced by one laser period, which are clearly observed on Fig.\ref{fig-PIC1D-Brunel} and Fig.\ref{fig_ejection}. This well-identified process is known as the Relativistic Oscillating Mirror (ROM) effect \cite{Lichters1996, article_NP, dromey_NP, gordienko2004relativistic,Baeva2006,Gonoskov2011,debayle2015self,Sanz2012,cherednychek2016analytical}. In simulations, the resulting periodic light emission has a spectrum consisting of a comb of high-order harmonics of the laser frequency: this is the origin of the harmonic signal observed in our experiment. The electron-harmonic correlations observed experimentally is therefore a signature of the periodicity of the dynamics of electrons.

As described in section \ref{Section:exp resu}, the harmonic signal is observed to collapse for longer density gradients $L$. One possible interpretation for this collapse can be that the electron emission ceases to be periodic in time.  And indeed, PIC simulations for longer density gradients (Fig.\ref{fig_ejection}d) strikingly show that, in contrast to the Brunel mechanism, electron emission is no longer periodic in this regime. The absence of harmonic signal in conjunction with the relativistic electron emission can thus be considered as a signature of the transition to a new coupling mechanism, associated to the very different plasma temporal dynamics observed in Fig.\ref{fig_ejection}. This mechanism will be described in section \ref{mechanism-long}. 

\subsection{Spatial structure of the electron emission} \label{e-spatial}

We now discuss the spatial properties of the outgoing electron beams, with the support of the simulation results of Fig.\ref{fig_ejection}. We show that in the short gradient regime, this structure is mostly determined by the interaction of expelled electrons with the reflected laser field in vacuum (section \ref{e-spatial-short}), while in the long gradient regime, it is rather imposed by large quasi-static surface fields that develop in the vicinity of the plasma surface during the interaction (section \ref{e-spatial-long}).

\subsubsection{Short-gradient regime}  \label{e-spatial-short}

In the case of short density gradients, the peculiar angular structure of the electron beam has recently been analyzed experimentally and theoretically in Ref. \cite{VLA}. In this Brunel regime, electrons are expelled from the plasma as a very laminar beam, with relativistic velocities initially quasi-parallel to the direction of specular reflection (Fig.\ref{fig_ejection}a). These relativistic electrons thus co-propagate with the intense reflected laser field, with which they interact in vacuum over a distance of the order of the Rayleigh length. This interaction always results in the ejection of electrons out of the laser beam,  and therefore digs a hole in the electron beam, centered on the specular direction, as observed in our experiment. 

There are two typical scenarios for this ejection, depending on the exact initial conditions with which electrons are expelled from the plasma into vacuum. Some electrons explore multiple optical cycles of the laser field, and thus oscillate in the field and get expelled from the laser focal volume by the so-called ponderomotive effect, isotropically and with a limited energy gain \cite{malka1997experimental, Quesnel}. They form the ring-shaped halo observed on the electron beam. But most electrons actually remain around a given phase of the reflected field and rather 'surf' a single wavefront of the reflected field, thus escaping the laser beam laterally along the laser polarization direction, and forming the bright peak observed next to the specular direction. The side on which this peak forms is determined by the laser phase at which electrons are expelled from the plasma into vacuum: the observation of a peak on one side only of the 'ponderomotive hole' (between the specular direction and the target normal) is a clear indication that electrons are ejected periodically \textit{once} every laser period, when the laser field drags them out of the plasma, in the form of a sub-optical cycle bunches. 

This second set of electrons experiences a quasi-constant $E$-field from the laser in vacuum until they escape the focal volume, leading to a greater energy gain than in ponderomotive scattering: this 'vacuum laser acceleration' (VLA) process accounts for the observed asymmetry of the $(\theta_x,E)$ distribution (see Fig.\ref{fig-Exp-Principle}b), where higher energies are observed on one side of the hole (VLA electrons) than on the other ('ponderomotive' electrons), as well as for the angle-energy correlations on this distribution \cite{thevenet2016physics}. An important consequence is that the Doppler upshift factor induced by outgoing electrons on the reflected field, which leads to the generation of high-order harmonic (ROM mechanism), cannot be directly deduced from the electron spectra measured experimentally, since electrons keep gaining energy after they escaped the target and emitted high-order harmonics, and before being detected. For instance, simulations show that in the present experiment, the electron Lorentz factor typically varies from $\gamma\approx 2-3$ as they are ejected from the plasma and emit harmonics, to $\gamma \gtrsim 15$ after their interaction with the reflected laser field \cite{VLA}, when they are detected.

\subsubsection{Long-gradient regime} \label{e-spatial-long}

The spatial properties of the electron beam observed in the 'long-gradient' regime described here have not been explained in detail yet, to the best of our knowledge. The electron trajectories displayed in Fig.\ref{fig_ejection}  show that the conditions of electron ejection from the plasma are already very different in the short and long gradient regimes. In the second case, the expelled electron beam is no longer laminar, and rather has a complex velocity distribution. As will be shown in the next section, this feature can be attributed to the chaotic character of the electron heating mechanism leading to ejection from the plasma. 

Furthermore, these 2D PIC simulations show that a quasi-static magnetic field develops at the plasma surface (see map of the magnetic field in Fig.\ref{fig_ejection}c, where a red area indicating a high magnetic field is present close to the plasma surface). This field grows during the laser-plasma interaction, reaches an amplitude typically of the same order of magnitude as the laser magnetic field, and then persists even after the laser pulse has been reflected by the plasma. A detailed analysis of electron trajectories in these simulations shows that this surface field, which is much larger than the one occurring in the short gradient regime (compare maps of the magnetic field in Fig.\ref{fig_ejection}a and c), strongly deflects the escaping electrons toward the target surface. This deflection accounts for the fact that the electron angular distributions observed in this regime are essentially centered between the specular direction and target surface. Test simulations have been performed to check that the reflected laser field plays no role in the deflection of electrons after their ejection from the plasma, in strong contrast to what is observed in Brunel regime.

Such surface quasi-static fields have already been reported in multiple studies of the interaction of intense lasers with dense plasmas (see e.g. \cite{Tatarakis, ruhl1999collimated, li2006observation, nakatsutsumi2018self}), and can be induced by a variety of physical processes (see e.g. \cite{fabbro1982hot,thaury2010influence,kumar2018energy,Sudan}). In the present case, our simulations indicate that their development can be attributed  to the 'fountain effect' described in \cite{Sudan} where they originate from the plasma cold return current that compensates for the lateral charge ejection from the laser focal volume. 
  This is supported by the fact that these fields are not observed at all in plane wave simulations, where the plasma surface is homogeneously illuminated by the laser field. 

In contrast to the short-gradient regime, here the spatial properties of the electron beam do not provide much insight into the involved electron heating mechanism. To identify this mechanism, we now turn to a more detailed numerical investigation based on PIC simulations.

\subsection{Electron heating mechanism in the long gradient regime} \label{mechanism-long}

\subsubsection{Importance of the reflected laser field} \label{comparison-1DPIC}

\begin{figure}[t]
\centering \includegraphics[width=\linewidth]{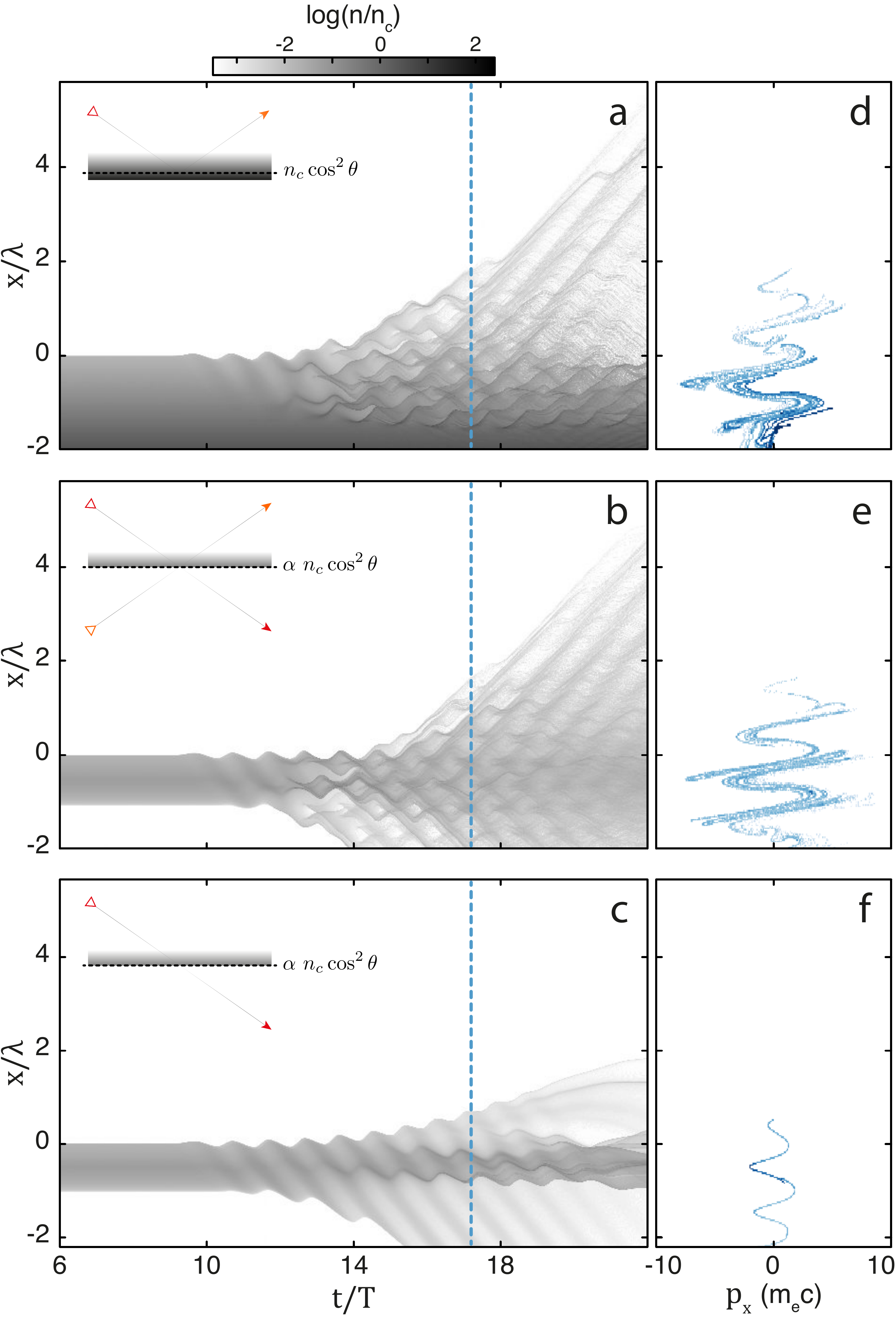}
\vskip -0.25cm 
\caption{\textbf{Set of 2D plane-wave PIC simulations carried out to reveal the role of the laser field reflected by the plasma}. Each line corresponds to a different physical case (see insets in panels a-c). case A (upper line) corresponds to an overdense plasma with a density gradient of scale length $L=\lambda$, irradiated by a single laser beam with an incidence angle $\theta_i=55^o$ and $a_0=2.5$. In case B, this plasma has been truncated for $n>0.4 n_c \cos^2 \theta$, keeping only the underdense part of the plasma, which is now irradiated by two almost identical laser beams (same parameters as in case A) arriving symmetrically from both sides of the plasma layer. In case C, this same plasma layer is irradiated only by the upper laser beam. In these three cases, plots of the temporal evolution of the plasma electron density (panels a-c) and snapshots of the $x-p_x$ phase space distribution of electrons (panels d-f, taken at the time indicated by the blue line in panels a-c) are displayed. The multilayered phase space distributions of panels d and e strongly contrast with the smooth regular distribution of panel f, and are typical of chaotic dynamics resulting from a repetitive stretching and folding effect in phase space.
}
\vskip -0.5cm
\label{fig-SH1}
\end{figure}

Our analysis of the electron heating mechanism in the long gradient regime is based on a set of 2D plane wave PIC simulations \cite{NotePIC} for three different physical configurations. Their key results are summarized in Fig.\ref{fig-SH1}, and shed light on the underlying physical mechanism. The upper panels (case A) correspond to the interaction of an ultraintense laser pulse with a dense plasma in the long gradient regime, i.e. the same physical configuration as in Fig.\ref{fig_ejection} and as in our experiment: they display the temporal evolution of the plasma electron density (panel a), and the $x-p_x$ phase space distribution of electrons (panel d) at the time when electron ejection from the plasma is observed to start (blue dashed line in panel a). 

The middle panels (case B) display the same quantities, but now in a situation where the plasma profile has been truncated for densities $n >  0.4 n_c \cos^2 \theta$, i.e. keeping only the underdense part of the plasma, such that there is hardly any reflection of the incident laser by the plasma. This underdense plasma layer, surrounded by vacuum on both sides, is irradiated by the same laser beam as before, but also by a second beam of slightly lower intensity (80\%), symmetrically arriving from the other side of the plasma. The role of this second laser is to emulate the beam reflected by the dense part of the plasma in case A. The key point here is that both the temporal dynamics of the plasma density profile and the electron phase-space distributions look very similar for cases A and B.

 By contrast, if the truncated plasma layer is irradiated by one laser beam only (case C, lower panels), the plasma dynamics becomes totally different. More specifically, while similar upward electron ejections are observed in cases A (corresponding to the electron signal observed in our experiment) and B, this electron emission is strongly reduced in case C: less electrons are emitted, and they have much weaker velocities. 

This toy-model study leads to two important conclusions. (i) The comparison of cases A and B indicates that in the long gradient regime, the coupling mechanism leading to electron ejection mostly occurs in the underdense part of the density gradient. (ii) The comparison of cases B and C indicates that the overdense part of the plasma nonetheless plays a key role, by producing a reflected beam that, when crossing and interfering with the incident beam, strongly modifies the dynamics of electrons in the underdense plasma layer. As we explain in the next section, the electron heating mechanism coming into play in such a physical situation is already well-identified in the existing literature. 

\subsubsection{Introduction to stochastic heating} \label{intro-SH}

Let us start by considering a \textit{free} electron (\textit{i.e.} without any collective plasma effects involved) exposed to two non-collinear ultraintense laser beams. Using a quantum description of the field as an ensemble of photons provides a simple way to understand that this electron can gain more energy than when exposed to a single laser beam. When a single laser beam (assumed to be a plane wave) is present, it is well-known that photon absorption processes are hindered because they do not enable to conserve both energy and momentum of the total system. By contrast, when two non-collinear beams are present, the combined absorption of multiple photons \textit{simultaneously from both beams} is allowed, because the availability of photons with different $\textbf{k}$ vectors makes it possible to conserve both energy and momentum of the total system. In other words, the presence of a second beam allows for energy absorption by the electron from the laser field, through a process that can be defined as stimulated multiphoton Compton scattering \cite{meyerhofer1997high}. 

For large field amplitudes, the laser field can be treated classically, and many previous studies in the literature have shown that electron dynamics in these combined noncollinear fields is not integrable and gets chaotic for high enough laser amplitudes (typically $a_0 \gtrsim 0.15$ for at least one of the two beams). This results in large energy gains, and this efficient regime of energy absorption by electrons is known as stochastic heating \cite{mendonca1983threshold,rax1992compton, ZMS2002, patin2006stochastic, bourdier2005stochastic} -although the name chaotic heating would probably be more appropriate here, since the system is perfectly deterministic and involves no stochastic process. 

This effect is obviously not restricted to isolated free electrons: it can equally occur for electrons in an underdense plasmas, leading to an energy absorption process where neither collisions nor collective plasma effects play any major role. Such a coupling of the plasma with multiple laser beams has been studied experimentally in \cite{zhang2003laser} by exposing an underdense plasma to two laser beams (like in case B of Fig.\ref{fig-SH1}). It is also known to play a role in electron injection in laser-driven plasma wakefield accelerators by the colliding pulse scheme \cite{faure2006controlled, rassou2014role}. 

To the best of our knowledge, Ref. \cite{SentokuAPB} was the first to point out that this mechanism should also come into play when an small underdense plasma layer is present at the surface of a dense plasma exposed to a single laser beam. In this case, the required second non-collinear laser beam results from the reflection of the single input beam by the dense plasma. Electrons in the underdense plasma are then exposed to the standing wave formed in front of the dense plasma by the superposition of the incident and reflected beams, and can gain energy by stochastic heating: this is precisely how we interpret our present experimental results in the 'long' gradient regime. 

\subsubsection{Numerical evidence for stochastic heating in the long gradient regime} \label{SH2}

With the help of simulations, we now support this interpretation by providing evidence that stochastic heating indeed occurs in the long gradient regime. To this end, we analyze the electron phase space distributions in cases A, B and C (panels d-f in Fig.\ref{fig-SH1}) and their temporal evolution during the laser-plasma interaction (see movie M1). When a single laser beam is present (case C), electrons are  observed to simply oscillate non-linearly in the laser field, leading to a smooth and regular phase space distribution (panel f and movie M1). In striking contrast, in cases A and B, electron dynamics in the standing wave resulting from the superposition of two non-collinear laser beams is complex: the key point is that we observe a very strong local 'stretching and folding' effect on the phase-space distribution, around each node of the standing wave (see movie M1). 

Such a stretching and folding effect results in very different trajectories for particles that are initially very close in phase space: this is known to be one of the most typical routes to chaotic dynamic \cite{strogatz2018nonlinear}, exemplified in the well-known horseshoe map models. These repetitive stretching and folding eventually result in a highly-structured, multilayered phase space distribution (panels d-e), where electrons at a given spatial position have a complex momentum distribution, typical of chaotic dynamics. The striking contrast between these highly-structured phase space distributions, and the smooth distribution observed in case C (Fig.\ref{fig-SH1}f) again demonstrates the impact of stochastic heating on electron dynamics in the underdense part of the plasma.  Furthermore, the comparison of these phase space distributions with the type of distribution observed for the Brunel mechanism (Fig.\ref{fig-PIC1D-Brunel}) is illustrative of the major difference in the dynamics of the system between the short and long gradient regimes. 

\begin{figure}[t]
\centering \includegraphics[width=\linewidth]{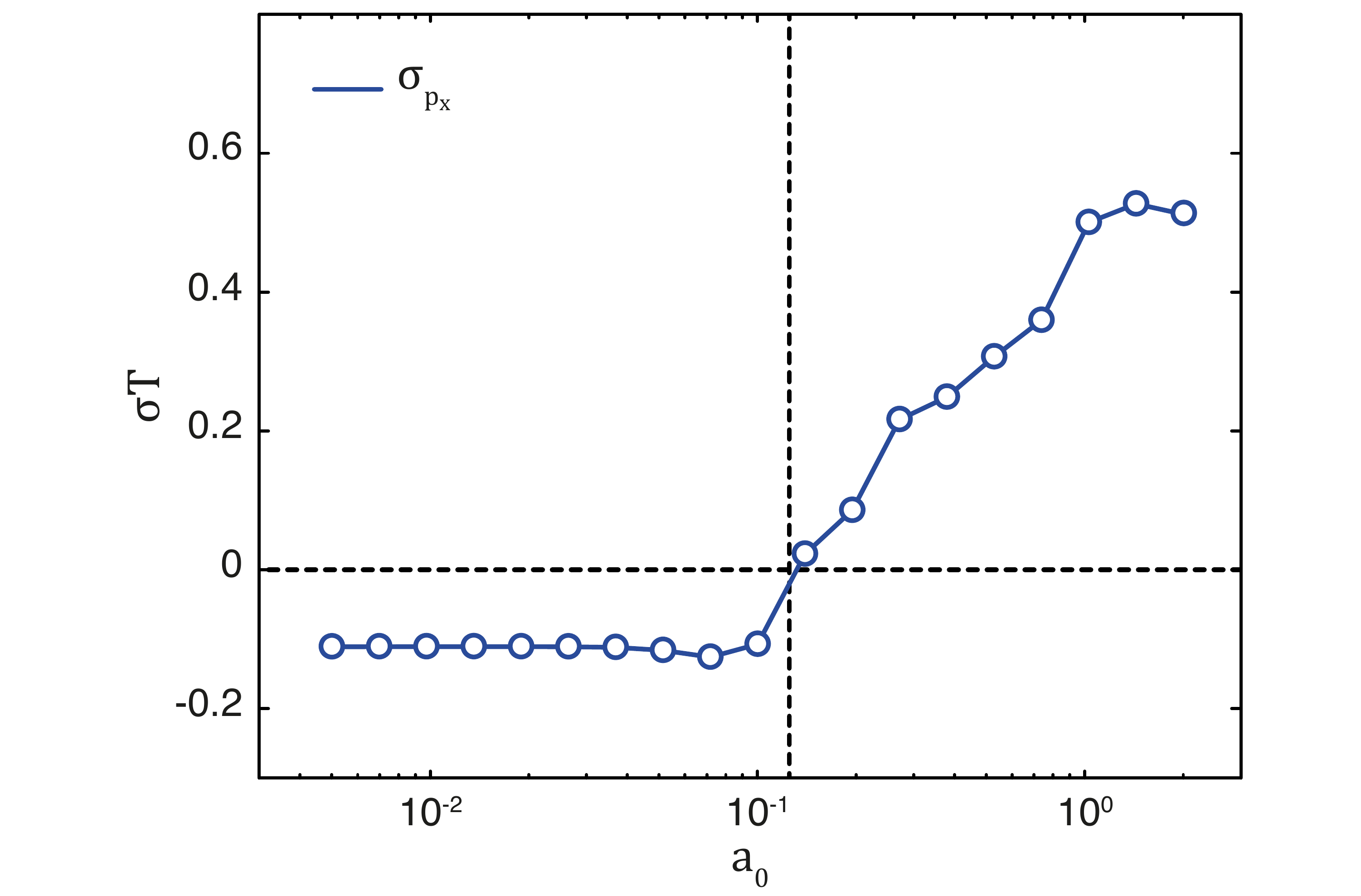}
\vskip -0.25cm 
\caption{\textbf{Lyapunov exponent for the $p_x$ variable in the long gradient regime, as a function of $a_0$.} Simulations have been performed in the physical configuration of case A of Fig \ref{fig-SH1}.  The onset of stochastic heating in the standing wave formed in the underdense part of the plasma by the incident and reflected laser fields is indicated by the fact that the Lyapunov exponent becomes positive. The calculation procedure used to deduce this exponent from the results of 2D plane-wave PIC simulations is explained in the supplementary material.   } 

\vskip -0.5cm
\label{fig-Lyapounov}
\end{figure}

The chaotic character of the electron dynamics can be further supported by the calculation of the Lyapunov exponents of plasma electrons (see supplementary material), which should be positive in the case of chaotic dynamics. The Lyapunov exponent $\sigma_{p_x}$ for the $p_x$ variable, obtained from 2D plane wave PIC simulations of case A of Fig.\ref{fig-SH1}, are displayed in Fig.\ref{fig-Lyapounov}, as a function of the incident laser amplitude. This exponent is negative at low intensity, and gets positive when $a_0 \gtrsim 0.15$, thus pointing to chaotic dynamics. This threshold in laser intensity is consistent with early theoretical investigations of stochastic heating \cite{mendonca1983threshold, ZMS2002}.

\subsection{Effect of the polarization, plasma reflectivity} \label{Energy}

With the support of the previous physical analysis, we finally discuss the experimental observations on the influence of the laser polarization direction (Fig.\ref{fig-Exp-PS}), and the evolution of the plasma reflectivity with $L$, for both  $s$ and $p$ polarizations (Fig.\ref{fig-Reflect}). 

In the short gradient regime, switching the polarization from $p$ to $s$ is expected to suppress both electron emission and harmonic generation, as observed experimentally (Fig.\ref{fig-Exp-PS}a-b), because the laser $E$-field component normal to the plasma surface is the main driving force of Brunel absorption at the laser intensities considered here. In contrast, when stochastic heating is involved, the polarization direction of the incident laser beam is not expected to have a strong influence, since the plasma surface only comes into play by producing a reflected wave: this explains why, in the long gradient regime, the electron signal is experimentally observed to still persist in $s$-polarization (Fig.\ref{fig-Exp-PS}c-d). This experimental observation alone makes a strong case against an interpretation of the long gradient regime in terms of resonant absorption, which should rather be very sensitive to the laser polarization. 

\begin{figure}[t]
\centering \includegraphics[width=\linewidth]{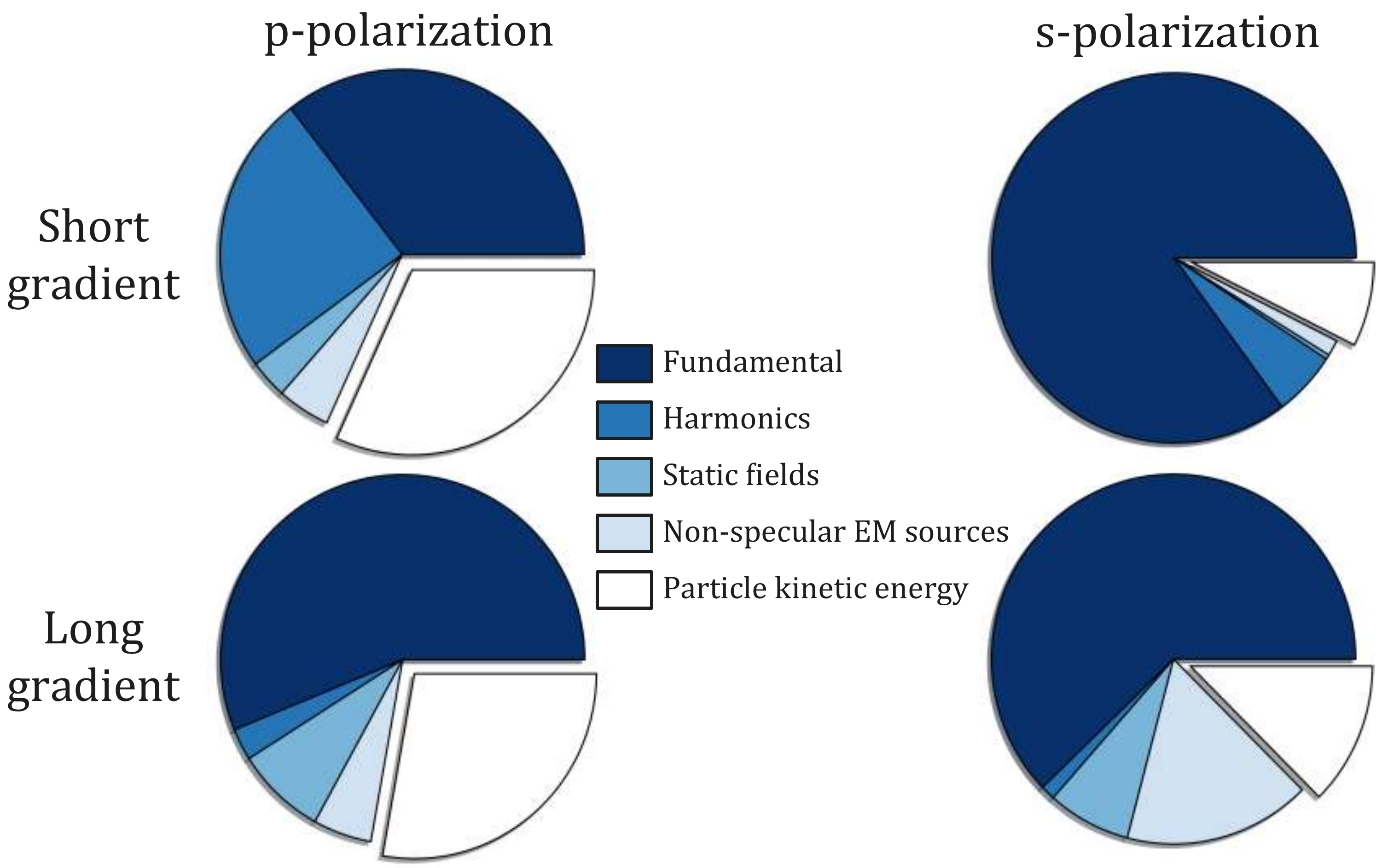}
\vskip -0.25cm 
\caption{\textbf{Distribution of the initial laser pulse energy after the interaction, in different conditions.} These pie charts summarize how the laser energy is distributed after the laser plasma interaction, in the short (upper line) and long (lower line) gradient regimes, for $p$ (left column) and $s$ (right column) polarizations. Five categories have been numerically separated, indicated in the legend: electromagnetic energy in the fundamental laser frequency in the specular direction, in the harmonics of the laser frequency in the specular direction, in quasi-static fields (i.e. with a frequency lower than the laser frequency), in non-static fields in the non-specular direction, and kinetic energy of plasma particles. The results are obtained from 2D PIC simulations with $a_0=3$, $\theta_i=55^o$, and $L_1=\lambda/15$ for the short gradient regime, $L_2=\lambda/1.5$ in the long gradient regime. } 

\vskip -0.5cm
\label{fig-Rep-Energie}
\end{figure}

In the long gradient regime, a spatial degradation of the reflected laser beam wavefronts is observed right after the target in these simulations (Fig.\ref{fig_ejection}c), while the beam wavefront is preserved in the short gradient regime (Fig.\ref{fig_ejection}a). This is qualitatively consistent with experimental observations (Fig.\ref{fig-Reflect}), where the laser beam intensity profile far from the target is observed to become degraded for long gradients. The chaotic character of the electron dynamics, identified in the previous section, affects the laser beam propagation in the underdense plasma layer, and thus provides a possible interpretation for this degradation of the reflected laser wavefronts.

From a more quantitative point of view, 2D PIC simulations also properly reproduce the evolution with $L$ of the plasma reflectivity at the fundamental laser frequency, for both $s$ and $p$ polarizations (Fig.\ref{fig-Reflect}). For short density gradients, the reflectivity is lower in $p$ ($\approx50 \%$) than in $s$ (close to $100 \%$) polarization, while similar values ($\approx 70 \%$) are observed for both polarizations in the long gradient regime. This is consistent with the results of Fig. \ref{fig-Exp-PS} on relativistic electron emission, and with our previous interpretation of the interaction: in $s$-polarization, for short $L$, Brunel absorption is suppressed, while for long $L$ stochastic heating is still present.

These 2D PIC simulations can be exploited to get more insight into the redistribution of the initial laser energy after the laser-plasma interaction. The different distributions obtained for both short and long gradient regimes, and for both $p$ and $s$ laser polarizations, are displayed as pie charts in Fig. \ref{fig-Rep-Energie}.  For short gradients and $p$-polarization, around 25 \% of the laser is converted into harmonics of the laser frequency (mostly low-orders), and about one third is deposited as kinetic energy of the plasma particles (among which the relativistic electrons observed in our experiment). When the polarization is switched to $s$, these two contributions get considerably reduced, down to around 5\% each, leading to a much higher reflectivity of the fundamental laser frequency. For long gradients, by contrast, very little energy is converted into harmonics, regardless of polarization. As already emphasized, the energy stored into quasi-static fields around the plasma surface significantly increases compared to the short gradient regime. In $p$-polarization, the fraction of energy going into particle kinetic energy is only slightly weaker than in the short gradient case, and gets reduced by about $50 \%$ in $s$-polarization. The reflectivity for the fundamental frequency remains similar for both polarizations, while more energy goes into scattered light in $s$-polarization.

\begin{figure*}[t]
\centering \includegraphics[width=\linewidth]{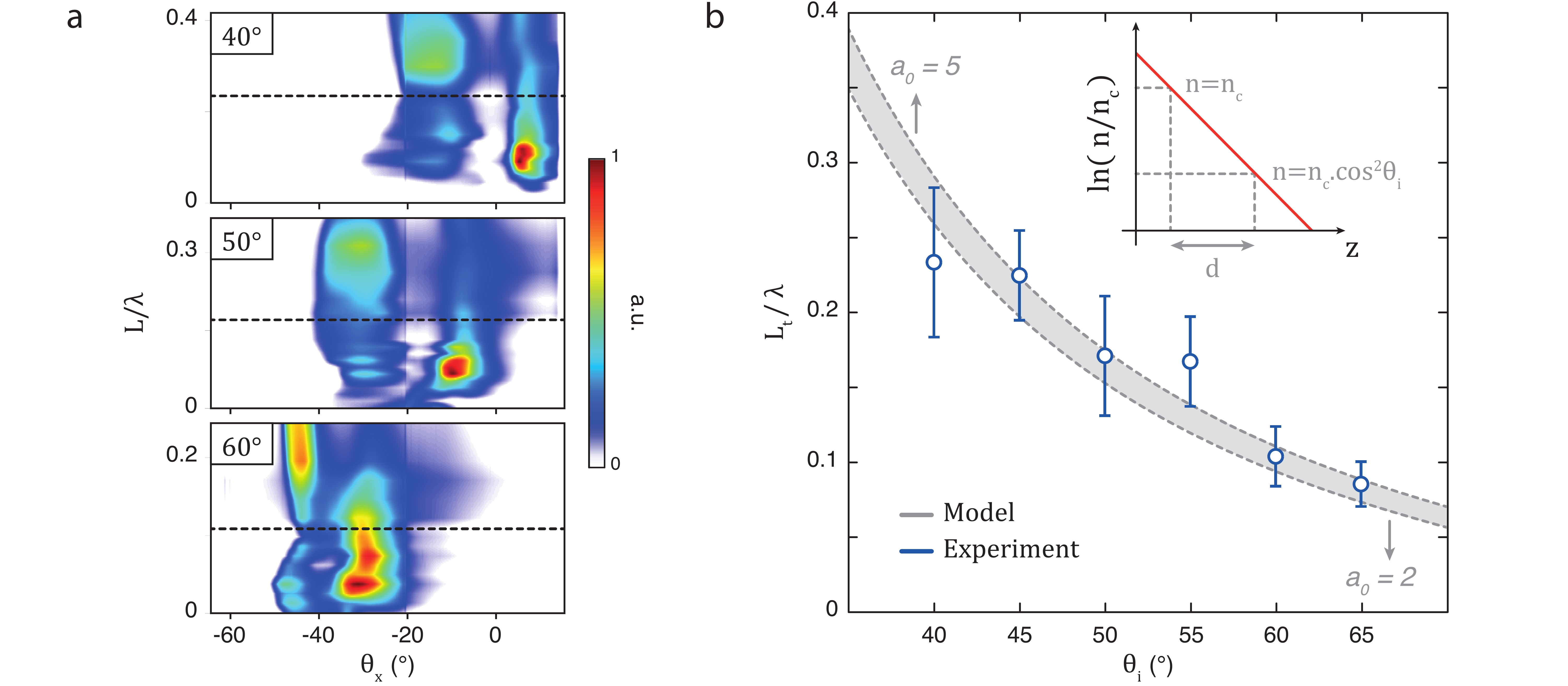}
\vskip -0.25cm 
\caption{\textbf{Effect of the laser incidence angle on the laser-plasma coupling.} Panel a shows the evolutions with $L$ of the high-energy electron beam angular profile in the incidence plane, measured for three different incidence angles, for a laser field amplitude $a_0\approx 3.5$. For each angle, $L$ is measured experimentally using the SDI technique. The value $L_t^e$ where a transition occurs is clearly observed to decrease with $\theta_i$. The measured evolution of $L_t^e$ with $\theta_i$ is plotted in panel b, where it is compared to the prediction of the simple model described in the text. The predictions of this model are shown for a range of laser amplitudes $2 \leq a_0 \leq 5$, to account for the experimental uncertainty on this amplitude. These predictions are very close to the high-amplitude limit given by Eq.(\ref{Lt}).  }

\vskip -0.5cm
\label{fig-Exp-Angle}
\end{figure*}

\subsection{Conclusion of the physical analysis}

The previous combination of multiple experimental observables and PIC simulations has provided strong evidence for the transition from Brunel absorption to stochastic heating as the density gradient scale length $L$ is increased, while no evidence of resonance absorption has been observed in this ultrahigh laser intensity regime. 

The intuitive physical insight underlying this transition is that the Brunel mechanism requires a sharp interface, such that the amplitude of the quivering motion of electrons in the laser field exceeds the length scale of this interface. On the opposite, stochastic heating can only occur if electrons are present within the standing wave interference pattern formed by the incident and reflected fields: it will therefore be favored by longer density gradients, for which this interference zone contains a larger number of particles provided by the underdense part of the plasma located just in front of the laser reflection point.

It however remains difficult to predict analytically the value $L_t$ of the density gradient scale length for which this transition occurs. In the next section, we determine this transition length experimentally as a function of the laser incidence angle on target $\theta_i$ - an essential physical parameter, since it affects the point of the density gradient where laser reflection occurs. We also discuss the effect of this angle on the properties of the emitted electron beams.

\section{Effect of the laser incidence angle}

\subsection{Evolution of the transition gradient scale length}

We have repeated the previous measurements, carried out for $\theta_i=55^o$, for five other incidence angles $\theta_i$ ranging from 40$^o$ to 65$^o$. For each angle, the density gradient scale length was systematically varied by changing the prepulse delay, and was measured using the SDI technique. The main outcomes of this experiment are summarized in Fig.\ref{fig-Exp-Angle}. The left panels show how the electron beam angular profile in the incidence plane evolves as a function of $L$, for three different incidence angles.  In all cases, the same type of transition as reported in Fig.\ref{fig-Exp-Principle} and \ref{fig-Exp-Resu} is observed, and it occurs for shorter values of $L$ as the incidence angle is increased. As before, we observe that this transition is correlated with major changes in the electron angle-energy distribution and in the harmonic emission. The experimental transition length $L_t^e$ deduced from all these data is plotted in Fig.\ref{fig-Exp-Angle}b (blue dots), and clearly decreases with $\theta_i$. 


To get some qualitative insight into this angular dependance, we consider the starting point of Brunel's model \cite{brunel}, which is that the physics of the laser-plasma coupling changes when the quivering amplitude $\Delta z_e$ of plasma electrons starts exceeding the typical spatial extent $d$ of the plasma-vacuum interface. We then define -somewhat arbitrarily- this spatial extent as $d=\left|z_s-z_c\right|$, the distance between the effective reflective surface of the plasma, located at $z_s(\theta_i)$ such that $n(z_s)=n_c \cos^2\theta_i$, and the fixed point corresponding to the location of the critical plasma density, $z_c$ such that $n(z_c)=n_c$ (see sketch in Fig.\ref{fig-Exp-Angle}b). $d$ can be calculated by using the standard assumption of an exponential density profile at the target surface \cite{zeldovich}: $n(z)=n_0 \exp(-z/L)$, with $n_0$ the maximum plasma density in the target, reached in $z=0$. This leads to $d=-2L \ln(\cos \theta_i)$.

In the simplest possible approach, the electron quivering amplitude can be considered to be the one of free electrons in the laser field, which is $\Delta z_e \approx \lambda /2 \pi \times a_0^2/(a_0^2+1)$.  At sufficiently high intensity, well into the relativistic regime ($a_0 \gg 1$), $\Delta z_e \approx \lambda /2 \pi$ is quasi-independent of the laser field amplitude. The condition $\Delta z_e=d(\theta_i)$ then leads to a simple theoretical expression for the transition gradient scale length (see movie M1):
\begin{equation} 
\label{Lt}
\frac{L_t}{\lambda}=- \frac{1}{4 \pi \ln(\cos \theta_i)}
\end{equation} 
This simple analysis indeed predicts an angular dependence of $L_t$ on $\theta_i$: physically, this is because for a given plasma density profile at the surface (i.e. a fixed $L$), the distance $d$ increases with $\theta_i$, due to the well-known angular dependence of the effective critical density, $n_s=n_c \cos^2 \theta_i$. Panel b in Fig.\ref{fig-Exp-Angle} displays a quantitative comparison of the experimental values $L_t^e$ with the prediction $L_t$ of this model, showing a good agreement. Despite its extreme simplicity, this model can thus account for the angular dependence observed experimentally. We note however that it becomes inappropriate for small angles -a regime that could not be investigated in our experiment for practical reasons- since it predicts that $L_t \longrightarrow \infty$ as $\theta_i \longrightarrow 0$. 

\begin{figure*}[t]
\centering \includegraphics[width=\linewidth]{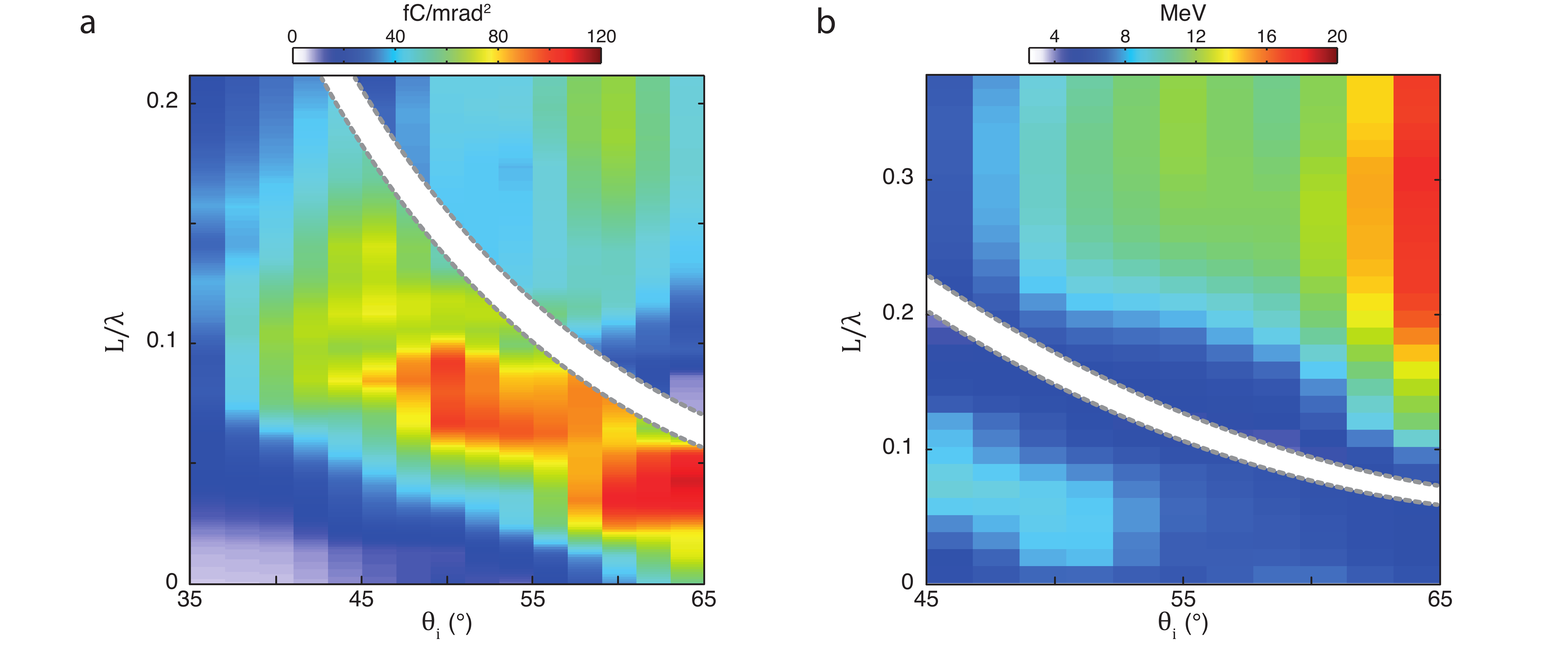}
\vskip -0.25cm 
\caption{\textbf{Measured electron beam properties as a function of key interaction parameters.} The left panel shows the experimental evolution of the number of emitted high-energy electrons ($E>1$ $MeV$) as a function of the incidence angle $\theta_i$ and density gradient scale length $L$, for $a_0=3.5$ and $p$-polarization of the laser. The gray area corresponds to the transition between the Brunel and stochatic heating coupling regimes, identified in Fig.\ref{fig-Exp-Angle}. Below this border, the displayed signal corresponds to the integration of the electron beam spatial profile on the right of the ponderometive hole (i.e. VLA electrons only). Above this border, it corresponds to the integration of the electron beam spatial profile on the left of this hole only. Panel b shows the average electron energy ($E_{m}=\int_{E>3 MeV} E\; n(E) dE /\int n(E) dE$) as a function of the same physical parameters. Here again, the electron spectra used for this calculation were selected on the right of the ponderomotive hole for short gradients, and on its left for long gradients.} 

\vskip -0.5cm
\label{fig-Exp-Angle2}
\end{figure*}

\subsection{Evolution of the electron number and energy}

Several other effects are observed when the incidence angle is changed, depending on the coupling mechanism. This is summarized in Fig.\ref{fig-Exp-Angle2}, where the evolutions of the electron signal and average energy are displayed as a function of the gradient scale length $L$ and incidence angle $\theta_i$, over a range that covers the short and long-gradient interaction regimes described before. In these plots, the areas associated to these two regimes are separated by a white zone, which corresponds to the transition curve predicted by the simple model of the previous section, and displayed in Fig.\ref{fig-Exp-Angle}b. In the following, we describe these evolutions and suggest tentative interpretations, which will need further investigations to be validated in detail. 

Both in the Brunel absorption and stochastic heating regimes, the number of emitted electrons grows with $\theta_i$ (see Fig.\ref{fig-Exp-Angle2}a) in the angular range investigated here. This might be attributed to the fact that when $\theta_i$ is increased, the target area covered by the laser focal spots increases, so that more electrons get involved in the interaction. 

As far as energy is concerned, it is hardly affected by the gradient scale length $L$ in each interaction regime, while the effect of incidence angle is different for the these two regimes. In the Brunel regime, the electron average energy only weakly changes with incidence angle, slightly decreasing for larger angles. Physically, in this regime, most of the electron energy gain is due to the VLA process, which is not expected to depend on the incidence angle since it occurs in vacuum. Changing $\theta_i$ might however modify the conditions of injection of electrons in the reflected field (e.g. their initial energy), and this might explain the observed slight angular dependence of the energy of VLA electrons.

By contrast, in the stochastic heating regime, the electron spectrum clearly shifts to higher energies as $\theta_i$ is increased (see Fig.\ref{fig-Exp-Angle2}b), up to about 20 $MeV$. This might be due to the fact that, all laser parameters being kept fixed, the 'life time' of the interference pattern formed by the incident and reflected fields increases with $\theta_i$. A simple analytical calculation indeed gives the following equation for the duration $\tau_{i}$ of this standing wave:
\begin{equation}
\tau_{i} = \sin \theta_i \left( \tau_L + \frac{w}{c} \: \tan \theta_i\right) 
\end{equation}
where $\tau_L$ is the laser pulse duration, and $w$ the laser focal spot size. As a result, for larger $\theta_i$, the stochastic heating process driven by the interference pattern can last longer, leading to a larger final energy gain for electrons, which could account for the experimental observation of Fig. \ref{fig-Exp-Angle2}.

\section{Conclusion}

In conclusion, we have combined state-of-the-art experiments and PIC simulations to provide the first unambiguous experimental evidence of the transition from Brunel absorption to a different laser-plasma coupling mechanism, as the steepness of the plasma surface is varied. This mechanism has been identified to be stochastic heating of electrons in the underdense plasma layer at the target surface, driven by the standing wave formed by the incident and reflected laser waves. This work has enabled the identification of clear signatures of these two coupling mechanisms, carried by the relativistic electron emission towards vacuum, the generated harmonic signal, and even the spatial profile of the reflected laser beam. At the laser intensities considered here, no evidence of the process known as resonant absorption has been found. 

These signatures should prove extremely useful for the interpretation of a broad range of topical experiments performed with high-power ultrashort lasers, related e.g. to ion and electron acceleration, or to the generation of short-wavelength light and attosecond pulses. In the later case, ultrahigh contrast pulses are required, generally obtained with plasma mirrors or by frequency-doubling after the final compression stage, and our results confirm that such experiments involve the Brunel mechanism. But in many other experiments -e.g. for ion acceleration from dense plasmas- preserving very steep interfaces is not strictly necessary. Such experiments are therefore often performed without contrast improvement after temporal compression: the laser pulses then typically have a very high temporal contrast up a few $ps$ before the main pulse, but their short-time contrast is often not as good, due e.g. to remaining high-order terms in the pulse spectral phase. With typical plasma expansion velocities in the 50 $nm/ps$ range, this so-called coherent pedestal on the $ps$ time scale will lead to density gradient scale lengths of the order of $\lambda$ at the arrival of the main pulse. In these conditions, stochastic heating should be the dominant coupling mechanism, and can now be readily be identified through the multiple signatures described in this work.

\begin{acknowledgments}

We are grateful to F. R\'eau, C. Pothier and D. Garzella for operating the UHI100 laser source. The research leading to this work has been funded by the ERC (grant ExCoMet, number 694596), the Agence Nationale pour la Recherche under contract ANR-14-CE32-0011-03 APERO, and LASERLAB-EUROPE (grant agreement no. 654148, EC's H2020 Framework Programme). 
An award of computer time (PICSSAR-INCITE) was provided by the Innovative and Novel Computational Impact on Theory and Experiment (INCITE) program. This research used resources of the Argonne Leadership Computing Facility (MIRA,THETA), which is a DOE Office of Science User Facility supported under Contract DE-AC02-06CH11357 as well as resources (CORI) of the National Energy Research Scientific Computing Center, a DOE Office of Science User Facility supported by the Office of Science of the U.S. Department of Energy under Contract No. DE-AC02-05CH11231.

\end{acknowledgments}

\bibliographystyle{h-physrev3}
\bibliography{biblio}

\begin{thebibliography}{10}
\bibitem{NotePIC}
When the incident laser beam has a finite focal spot size, these simulations
  are performed in the laboratory frame. When the incident laser beam is
  considered to be a plane wave (infinite focal spot size), the simulation is
  performed in a Lorentz-boosted frame, such that the laser impinges the plasma
  at normal incidence. Then, only one spatial coordinate (along the target
  normal) is taken into account in the simulation, but the three velocity
  components of particles are calculated. We refer to this second type of
  simulations as '2D plane wave'.
\end{thebibliography}

\end{document}